\documentclass[preprint,12pt]{elsarticle}

\usepackage{amssymb}
\usepackage{graphicx}% Include figure files
\usepackage{dcolumn}% Align table columns on decimal point
\usepackage{bm}% bold math
\usepackage[utf8]{inputenc}
\usepackage[T1]{fontenc}
\usepackage{mathptmx}
\usepackage{etoolbox}
\usepackage{amssymb}
\usepackage{siunitx}
\usepackage{amsmath}
\usepackage{esint}
\usepackage{float}
\usepackage{caption,subcaption}
\usepackage{csquotes}
\usepackage{booktabs}
\usepackage{hyperref}
\usepackage{multirow}
\usepackage[utf8]{inputenc}
\usepackage{textcomp}

\usepackage{graphicx}
\usepackage[version=4]{mhchem}
\usepackage{longtable,tabularx}
\usepackage{xcolor}
\usepackage{indentfirst}

\journal{arXiv}

\begin{document}
\begin{frontmatter}



\title{Performance analysis for a rotary compressor at high speed: experimental study and mathematical modeling}


\author[inst1,inst2]{Chuntai Zheng}
\author[inst2,inst3]{Wei Zhao}
\author[inst1]{Benshuai Lyu \corref{cor1}}
\ead{b.lyu@pku.edu.cn}
\cortext[cor1]{Corresponding author}

\author[inst2]{Keke Gao \corref{cor2}}
\ead{gaokk@midea.com}
\cortext[cor2]{Corresponding author}

\author[inst2]{Hongjun Cao}
\author[inst2]{Lei Zhong}
\author[inst2]{Yi Gao}
\author[inst2]{Ren Liao}

\affiliation[inst1]{organization={State Key Laboratory of Turbulence and Complex Systems, College of Engineering, Peking University},%Department and Organization
            addressline={5 Yiheyuan Road, Haidian District}, 
            city={Beijing},
            postcode={100871}, 
            % state={State One},
            country={China}} 

\affiliation[inst2]{organization={Midea Corporate Research Center},%Department and Organization
            addressline={Shunde}, 
            city={Foshan, Guangdong Province},
            postcode={528300}, 
            % state={State One},
            country={China}}

\affiliation[inst3]{organization={School of Mechanical Engineering, Shanghai Jiao Tong University},%Department and Organization
            addressline={}, 
            city={Shanghai},
            postcode={200240}, 
            % state={State One},
            country={China}}

% \author[inst2]{Author Two}
% \author[inst1,inst2]{Author Three}

% \affiliation[inst2]{organization={Department Two},%Department and Organization
%             addressline={Address Two}, 
%             city={City Two},
%             % postcode={22222}, 
%             % state={State Two},
%             country={Country Two}}

\begin{abstract}
This paper conducted a comprehensive study on the performance of a rotary compressor over a rotational speed range of 80Hz to 200Hz through experimental tests and mathematical modeling. A compressor performance test rig was designed to conduct the performance tests, with fast-response pressure sensors and displacement sensors capturing the P-V diagram and dynamic motion of the moving components. Results show that the compressor efficiency degrades at high speeds due to the dominant loss factors of leakage and discharge power loss. Supercharging effects become significant at speeds above 160Hz, and its net effects reduce the compressor efficiency, especially at high speeds. This study identifies and analyzes the loss factors on the mass flow rate and power consumption based on experimental data, and hypothesizes possible mechanisms for each loss factor, which can aid in the design of a high-speed rotary compressor with higher efficiency. 
\end{abstract}

\begin{graphicalabstract}
\includegraphics[width=0.6\textwidth]{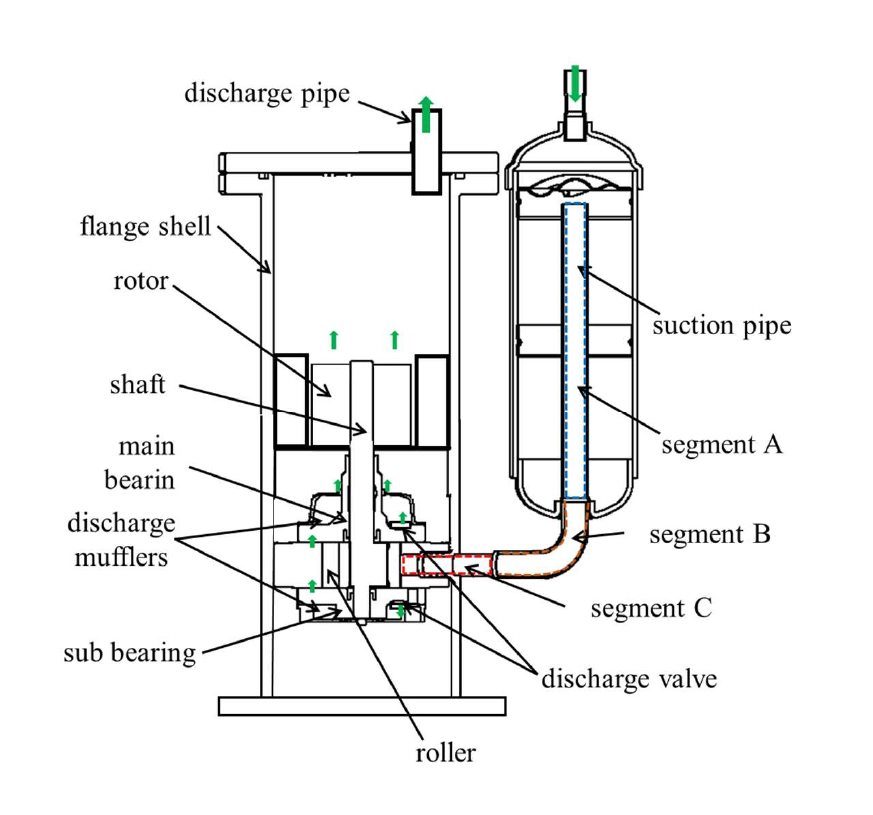}
\end{graphicalabstract}

\begin{highlights}
\item The performance of a rotary compressor was analyzed for high rotational speeds of up to 200Hz.
\item The loss factors were identified through a combination of experiment and mathematical modeling.
\item The net impact of each loss factor on the compressor efficiency was determined and quantified.
\end{highlights}


\end{frontmatter}
%\linenumbers

\section{Introduction}

Rotary compressors (rolling piston compressors), due to their advantages of simple structure, high efficiency, and fewer components, are widely used in household air conditioning and other fields, and their power consumption accounts for 85\% of the total energy consumption of air conditioning. In recent years, with the demand for energy conservation, emission reduction, and industrial upgrading, developing high-efficiency and energy-saving rotary compressors is of great significance.

The rotary compressor is a volumetric type compressor, mainly composed of a cylinder, roller, discharge valve, vane, crankshaft, and motor. The principle of the rotary compressor is to use an eccentric cylindrical roller to rotate in the cylinder to change the working volume of the cylinder, hence realizing the suction, compression, and discharge of the refrigerant.
Current research indicates that the high-speed, small-capacity compressor design helps to improve efficiency under near-real operating conditions. In addition, a high-speed design leads to smaller product sizes and therefore reduced manufacturing and transportation costs.

The rotary compressor was developed in the 1960s. After decades of research, the energy efficiency of the rotary compressor has gradually improved, and the design of components has been simplified significantly\cite{aw2021review}. Research on rotary compressors generally falls into three categories based on methodologies, i.e., experiment, numerical simulation, and mathematical modeling. 

Experimental tests of rotary compressors under various operating conditions are typically conducted through a calorimetric system. The energy efficiency of compressors can be evaluated by measuring the cooling capacity and power consumption of rotary compressors and analyzing the thermodynamic processes of refrigerants\cite{wu2015experimental}. Measuring the instantaneous pressure within the cylinder is an important approach to analyzing the power loss while analyzing the P-V diagram is an inevitable step of developing a new compressor\cite{wakabayashi1982analysis,nagatomo1984performance,monasry2018development,matsuzaka1982rolling,liu1994performance,youn1998design}. 

For instance, \citet{youn1998design} carried out P-V analyses to study the influence of various design parameters of the discharge system on power loss in the compression and discharge process. \citet{lee2012performance} measured the P-V diagram of a newly designed compressor to evaluate and improve the design. \citet{monasry2018development} optimized cylinder dimensions and designed a new discharge structure to improve compressor efficiency, and the effects were compared on the P-V diagram. Furthermore, to understand the factors that influence compressor performance, numerous attempts have been made to develop analytical models based on experimental data. \citet{wakabayashi1982analysis} proposed heat exchange equations based on test data to estimate leakage loss and heat transfer loss. \citet{liu1994performance} used the P-V diagram to study the net effect of supercharging on compressor performance. \citet{nomura1984efficiency} and \citet{nagatomo1984performance} presented several sets of gas equations based on the experimental data to assess the reduction in volumetric efficiency and identify power loss.     

With the advancement of computer technology, computational fluid dynamics (CFD) has become the primary means for rotary compressor design, involving fluid-solid coupling and heat transfer\cite{kim2017estimation}. For example, CFD simulations can be used to study the three-dimensional gas dynamics during the suction and compressor process\cite{ba2016gas}. The oil supply system under different operating conditions can also be analyzed using CFD simulation\cite{wu2013numerical}. However, CFD simulations are sensitive to numerical errors and must be validated against experimental data. Additionally, many real physical processes are simplified or ignored in CFD simulations, making it difficult to fully comprehend the physical phenomenon of rotary compressors. Through extensive experimentation and numerical simulations, researchers have developed mathematical models for the entire rotary compressor\cite{cipollone2006theoretical,wu2000mathematical,mathison2008modeling,cai2015simulation}. These models primarily consist of a thermodynamic model of refrigerant gas based on the principles of energy and mass conservation law\cite{padhy1994heat}, a heat transfer model to predict the transient heat transfer of the entire compressor system\cite{liu1992modeling}, unsteady models to predict the suction and discharge process, and models for calculating the oil and refrigerant mixture leakage flow\cite{liu1994performance,wu2000mathematical}. The existing mathematical models have been shown to accurately predict compressor performance, and the primary loss factors have been identified. 

Despite the extensive research on rotary compressors, our understanding of their performance remains incomplete. First, at rotational speeds (revolution per second) above 140Hz, there is a significant efficiency decay. Because most research on rotary compressors focused on speeds below 140Hz, the physical mechanism of this efficiency decay remains unclear. Second, existing mathematical models cannot accurately predict the efficiency decay at high speeds. Third, the majority of performance analysis relies on either mathematical modeling or experimental data without which a comprehensive analysis of efficiency losses is still difficult. 

This study aims to bridge the gap by studying the performance of rotary compressors at high speeds. First, experimental methods and mathematical modeling of losses are introduced. Second, experimental results with performance criteria are analyzed, and loss analysis is established to reveal the mass flow loss and power loss at high speeds. Third, discussions are made considering the net effects of loss factors on compressor performance, with special attention paid to the supercharging effects at high speeds. The remainder of this paper is organized as follows: Section \ref{sec:exp} introduces the experimental setup, Section \ref{sec:Lossanalysis} presents the mathematical modeling of loss factors, Section \ref{sec:results} shows the results and analysis of mass rate losses and power losses, revealing the significance of each loss factor at high speeds. Conclusions are drawn in Section \ref{sec:conclusion}.

\section{Experimental setup}\label{sec:exp}

\subsection{Compressor}
\begin{figure}[ht!]
  \centering
  \includegraphics[width=0.7\textwidth]{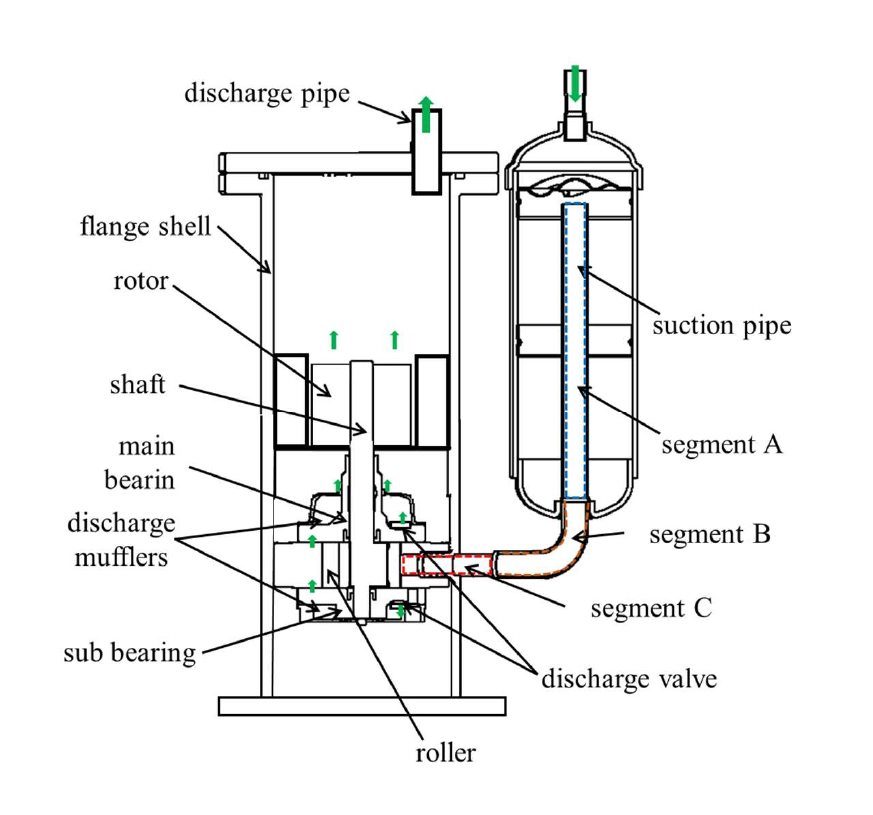}
  \caption{Construction of the rotary compressor (the green arrow indicates flow directions).}
  \label{fig:compressor}
\end{figure}

As shown in figure \ref{fig:compressor}, a hermetic-type rotary compressor was built for the test. The compressor consisted of a compression mechanism and a motor, both installed within a flange shell. The compression mechanism, consisting of a cylinder, a roller, and discharge valves, was placed in the lower section, while the motor was placed in the upper section. The cylinder has two discharge ports, one located at the main bearing and the other at the sub-bearing. As indicated by the arrow in figure \ref{fig:compressor}, the discharged refrigerant gas flowed through a vertical channel on the cylinder to mix with the gas discharged inside the upper muffler, after being discharged into the lower muffler. An appropriate amount of lubricating oil was sealed in the bottom section. The roller was mounted on an eccentric shaft and rotated around the cylinder center to generate the periodic volume change of the suction and compression chambers. A suction pipe with a reservoir was inserted into the suction chamber and the compressed gas in the compression chamber was discharged through the discharge valve into the flange shell. 

\subsection{Pressure and motion sensors}

\begin{figure}[ht!]
  \centering
  \begin{subfigure}[b]{0.49\textwidth}
    \includegraphics[width=\textwidth]{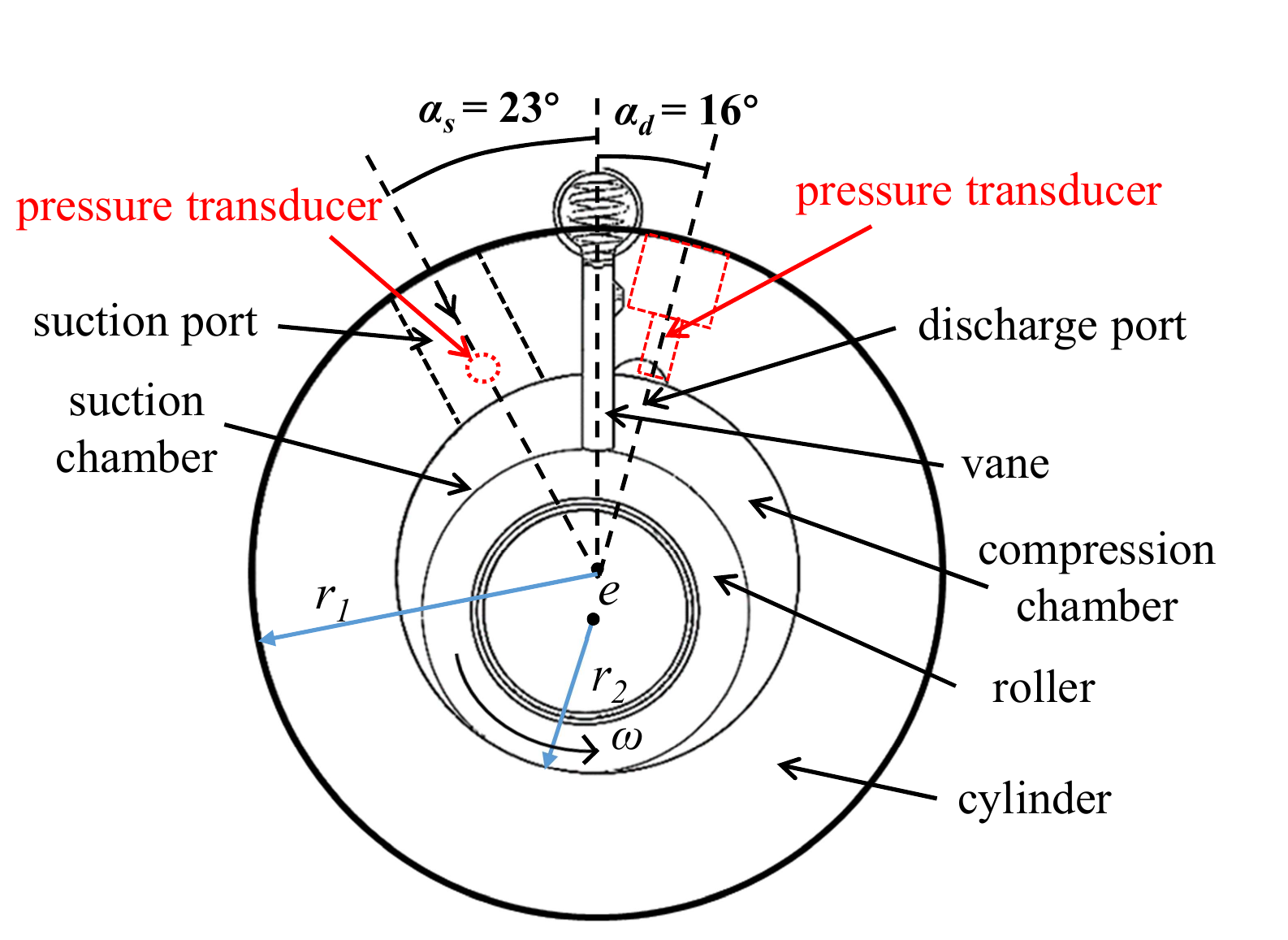}
    \caption{}
  \end{subfigure}             
  \begin{subfigure}[b]{0.49\textwidth}
    \includegraphics[width=\textwidth]{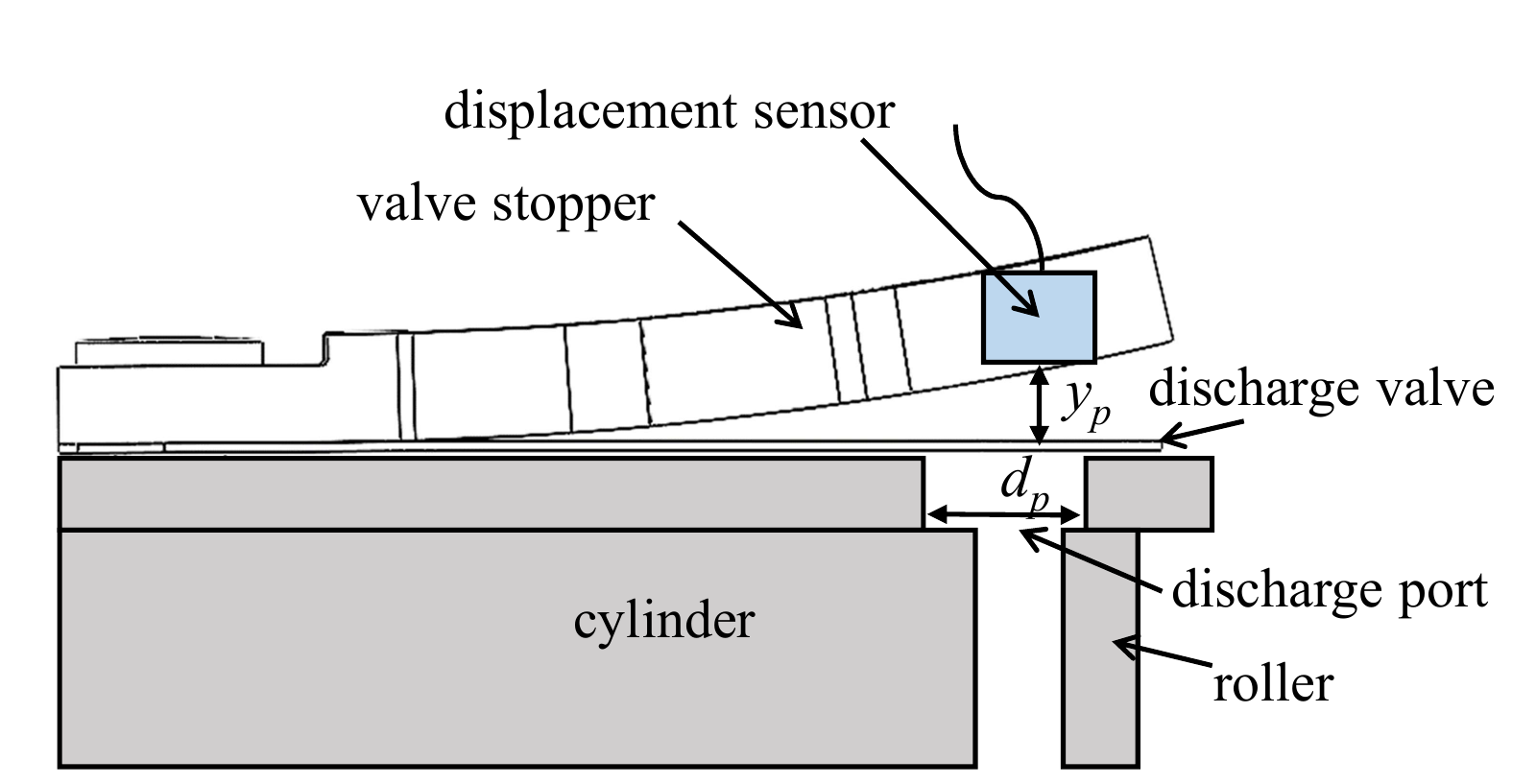}
    \caption{}
  \end{subfigure}             
  \caption{Configuration of (a) the compression mechanism and the installation location of the pressure transducer and (b) the discharge system and the installation location of the displacement sensor. }
  \label{fig:sensor}
\end{figure}

Two Endevco 8530B piezoresistive pressure transducers were used to measure the absolute pressure inside the chambers. The transducer has an uncertainty of $\pm$ 0.1\% within its dynamic range. The sensitivity of the pressure transducers was calibrated before the measurement. As shown in figure \ref{fig:sensor} (a), one pressure transducer was installed on the suction port, positioned at 23$^{\circ}$ to the center line of the vane groove. Another pressure sensor was installed on the side wall of the cylinder close to the discharge port, forming a 16$^{\circ}$ angle to the center line of the vane groove. 

As shown in figure \ref{fig:sensor} (b), two AEC displacement sensors were installed to measure the displacement of the sliding vane and the discharge valve. The sensor has a resolution of 2 $\mu$m working at a temperature range of $\SI{-20}{\degreeCelsius}$ to $\SI{120}{\degreeCelsius}$. One sensor was installed to measure the vane displacement. The minimum displacement of the sliding vane represents the zero shaft angle $\alpha = 0$ where the roller surface is tangent to the cylinder surface at the center line of the vane groove. Another sensor was installed on the valve stopper to measure the valve lift $y_p$. 

\subsection{Compressor test rig and test conditions}

\begin{figure}[ht!]
  \centering
  \includegraphics[width=0.7\textwidth]{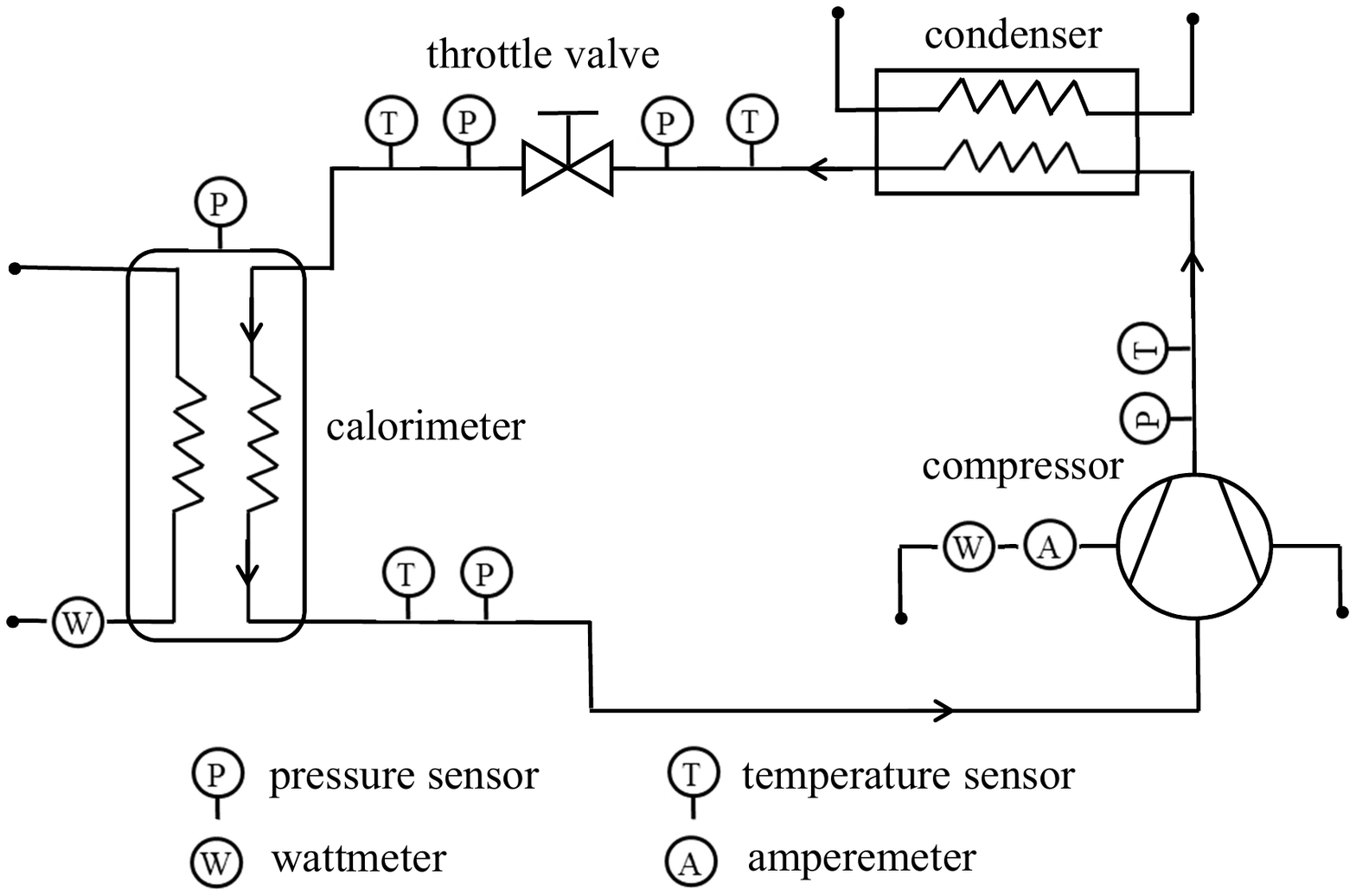}
  \caption{Schematic of the compressor performance test rig.}
  \label{fig:rig}
\end{figure}

As shown in figure \ref{fig:rig}, the tests were conducted at a 1kW $\sim$ 20kW compressor performance test rig at Midea Corporate Research Center. The test rig mainly includes a condenser, an evaporator, a calorimeter, a throttle valve, and some instruments or meters. The compressor was connected to the test rig by the suction and discharge pipes. The test rig was used to precisely control the operating conditions and measure the cooling capacity and input power of compressors. It has an accuracy of $\pm$0.005 Mpa for the pressure control and an accuracy of $\pm\SI{0.2}{\degreeCelsius}$ for the temperature control. The tests were carried out under identical conditions with varying rotational speeds, as listed in table \ref{tab:operation}. Note that the reference conditions refer to that of the refrigerant at the suction and discharge conditions, which are listed in table \ref{tab:operation}. 

\begin{table}[h]
  \caption{Test conditions}
  \begin{center}
  \begin{tabular}{cc}
  \toprule[2pt]
      Refrigerant ~  & Rotational speed (Hz)  \\[4pt]
      R32 ~~~  & 80;120;140;160;180;200\\
      \midrule[1pt]            
     Suction pressure (Mpa)  ~ &  Discharge pressure(Mpa) \\
     0.5 ~  & 2.1 \\ 
      \midrule[1pt]
     Condensing temperature (\textcelsius) ~  & Evaporating temperature (\textcelsius) \\ 
      33.3 & ~~~ -14.3\\ 
      \midrule[1pt]
      Suction temperature (\textcelsius) & ~~~ Temperature before valve (\textcelsius)\\ 
      -6 ~~~  & 25\\ 
      \bottomrule[2pt]
  \end{tabular}
  \label{tab:operation}
  \end{center}
\end{table}

\section{Performance criteria}\label{sec:performance}

The experimental data can be used to calculate the performance criteria of the compressor at various speeds. In this study, the total efficiency $\eta$ (compressor efficiency) is used to evaluate the overall performance\cite{liu1993simulation}, which can be defined as

\begin{equation} \label{eq:effi}
\eta = \eta_{v} \eta_{c} \eta_{f} \eta_{m}
\end{equation}

The rest of the terms in the equation \ref{eq:effi} are defined as:
\begin{equation}\label{eq1}
\begin{split}
        &\eta_{v} = \text{volumetric efficiency} = \frac{\text{actually discharged gas volume}}{\text{ideal discharged gas volume}} = \frac{V_2^{\prime}}{V_{2}} \,, \\
        &\eta_{c} = \text{cylinder process efficiency} = \frac{\text{ideal gas compression work}}{\text{actual gas compression work}} = \frac{W_{\text{ideal}}}{W_{\text{real}}} \,, \\
        &\eta_{f} = \text{mechanical efficiency} = \frac{\text{actual gas compression work}}{\text{motor shaft output}} =  \frac{W_{\text{real}}}{W_{\text{shaft}}}  \,, \\
        &\eta_{m} = \text{motor efficiency} = \frac{\text{motor shaft ouput}}{\text{motor electrical input}} =  \frac{W_{\text{shaft}}}{W_{\text{in}}}  \,,
\end{split} 
\end{equation}
where $V_2^{\prime}$ and $V_2$ are the actual and ideal discharged volume under the suction reference condition, respectively. $\eta_f$ is related to the friction losses in the compressor. A combined motor and friction loss coefficient $\eta_{mf}$ can be introduced by $\eta_{mf} = \eta_{m} \eta_{f} = W_{\text{real}}/W_{\text{in}}$, which is used to quantify the motor loss and the mechanical loss as a whole.

\section{loss analysis}\label{sec:Lossanalysis}

The total losses of the compressor operating at various speeds can be classified as mass gain or loss and power losses.  

\subsection{Mass flow gain and loss}\label{sec:massLoss}

Mass flow loss is defined by the difference between the theoretically discharged mass rate and the actual mass rate which can be measured by the compressor test rig. The loss factors can be classified as the suction gas heating loss, leakages, clearance volume loss, backflow loss, and supercharging effect. Each factor can be estimated by mathematical modeling based on the test data. When presenting the mass flow losses in relation to the volumetric efficiency $\eta_v$, the sum of the losses can be expressed as

\begin{equation}\label{eq:VEcal}
    \eta_v = 1 - \lambda_{sh} - \lambda_{lk} - \lambda_{cv} - \lambda_{bf} - \lambda_{sc} \,,
\end{equation}
where $\lambda_{sh}$ is the suction gas heating loss coefficient, $\lambda_{lk}$ is the leakage loss coefficient, $\lambda_{cv}$ is the clearance volume loss coefficient, $\lambda_{bf}$ is the backflow loss coefficient, and $\lambda_{sc}$ is the supercharging loss coefficient.

\subsubsection{Suction gas heating loss}

During the process of refrigerant gas entering the suction chamber, the gas continuously draws heat from the suction pipe and the cylinder wall. During the heating process, the gas expands resulting in lower density which leads to less mass pumped into the suction chamber. The effect of speeds on heat transfer is complex and requires evaluation through a heat transfer model. Higher rotational speeds can lead to less heat transfer time, but faster flow caused by increased rotational speed can also increase heat transfer coefficients. Therefore, the net effect of shaft speed on heat transfer needs to be carefully evaluated.

\paragraph{Heat transfer in the suction pipe}

The suction pipe has a circular cross-section. The convection heat transfer coefficient for turbulent flow in a circular pipe can be expressed as\cite{mills1992heat}
\begin{equation}
h_p=0.023 \frac{k}{D} {Re}_D^{0.8} {Pr}^{0.4} \,,
\end{equation}
where $k$ is the thermal conductivity of the refrigerant gas inside the pipe, $D$ is the pipe diameter, $Re_{D}$ is the Reynolds number based on the pipe diameter, and $Pr$ is the Prandtl number of the gas.

The whole pipe is divided into three segments as shown in figure \ref{fig:compressor}. The temperature of each segment was measured by attached thermocouples. By applying the energy conservation law to each segment, we have

\begin{equation}
\begin{split}
& h_p \pi D L_A\left(T_{A}-\frac{T_s+T_{s1}}{2}\right)=\dot{m}_{s} c_p\left(T_{s1}-T_s\right) \,,\\
& h_p \pi D L_B\left(T_{B}-\frac{T_{s1}+T_{s2}}{2}\right)=\dot{m}_{s} c_p\left(T_{s2}-T_{s1}\right) \,,\\
& h_p \pi D L_C\left(T_{C}-\frac{T_{s2}+T_{s3}}{2}\right)=\dot{m}_{s} c_p\left(T_{s3}-T_{s2}\right) \,,\\
\end{split}
\end{equation}
where $c_p$ is the specific heat at constant pressure, $\dot{m}_{s}$ is the mass flow rate, $T_s$ is the temperature of the gas entering the suction pipe, $T_{s1}$, $T_{s2}$, and $T_{s3}$ are the temperature of the gas leaving segment A, B, and C, respectively, $T_{A}$, $T_{B}$, and $T_{C}$ are the measured wall temperature of segment A, B, and C, respectively, and $L_{A}, L_{B}$, and $L_{C}$ are the length of the segment A, B, and C.

% Although $c_{p}$ and $h_p$ differ according to the changes in gas properties at different temperatures, they are calculated at $T_s$, as the temperature changes inside the suction pipe are not significant.

\paragraph{Heat transfer in the cylinder}

To calculate the heat transfer between the refrigerant gas and the cylinder wall, the following equation is used to predict the convection heat transfer $h_c$ inside the cylinder\cite{padhy1994heat}:

\begin{equation}
h_c=0.025 \frac{k}{D_h} {Re_{D_h}}^{0.8} {Pr}^{0.4}\left(1.0+1.77 \frac{D_h}{r_{\text {ave }}}\right) \,,
\end{equation}
where $r_{\text{ave}}$ is the average of $r_1$ and $r_2$ ($r_1$ is the radius of the cylinder and $r_2$ is the radius of roller), and $D_h$ is the hydraulic diameter expressed as $D_h = 4V(\alpha)/A_c(\alpha)$ where $A_c(\alpha)$ is the total heating area inside the cylinder, and ${Re_{D_h}}$ is the Reynolds number based on $D_h$ and characteristic velocity $u$ which can be expressed as $u = 2 \omega r_1$.

The suction gas inside the cylinder is heated by the cylinder inner wall $A_o$, rolling piston outer wall $A_p$, upper and lower cylinder head walls $A_h$, and vane wall $A_v$ \cite{1984741}. Thus, $A_c$ can be expressed as

\begin{equation}
    A_c(\alpha) = A_o(\alpha) + A_p(\alpha) + A_h(\alpha) + A_v(\alpha)
\end{equation}

The temperatures of the rolling piston outer wall and vane wall are assumed to be equal to the temperature of the lubricating oil $T_{os}$ which is measured by a thermocouple during the test. The temperature distributions of the cylinder inner wall $T_{c}(\alpha)$ can be estimated by the relationship\citep{1984741}:

\begin{equation}
T_{c}(\alpha)=\left(T_{o s}-3.5\right)+ \frac{2(\alpha-\pi)}{2 \pi}-0.0028\left(T_0-308.15\right)^2+0.178\left(T_0-308.15\right) \,,
\end{equation}
where $T_{0}$ is the suction temperature at inlet.

Therefore, the heat transfer rate between the cylinder and the gas can be expressed as

\begin{equation}
\begin{aligned}
\frac{d Q}{d \alpha} =& \frac{1}{\omega} h_c (A_o(\alpha)\left(T_{c}(\alpha)-T(\alpha)\right) + A_p(\alpha)\left(T_{os}-T(\alpha)\right) \\
                     &+ A_h(\alpha)\left(T_{c}(\alpha)-T(\alpha)\right) + A_v(\alpha)\left(T_{os}(\alpha)-T(\alpha)\right))  \,,\\
\end{aligned}
\end{equation}
where $T(\alpha)$ is the gas temperature.

By applying the first law of thermodynamics to the suction chamber as a control volume for an open system, the change of the internal energy $U$ can be expressed as 

\begin{equation}\label{eq:energy}
\frac{d U}{d \alpha}=\frac{d Q}{d \alpha}+\frac{d W_{\text{net}}}{d \alpha}+h_i \frac{d m_i}{d \alpha}-h_o \frac{d m_o}{d \alpha} \,,
\end{equation}
where $W_{\text{net}}$ is the work done to the control volume by the roller, $h_i$ is the specific enthalpy of the gas entering the control volume, $m_i$ is the mass entering the control volume, $h_o$ is the specific enthalpy of the gas exiting the control volume, and $m_o$ is the mass exiting the control volume.

The mass inside the control volume $m_c$ can be calculated by applying the mass conservation law:

\begin{equation}\label{eq:mass}
\frac{d m_c}{d \alpha}=\frac{d m_i}{d \alpha}-\frac{d m_o}{d \alpha}
\end{equation}

\begin{figure}[h!]
  \centering
  \includegraphics[width=0.5\textwidth]{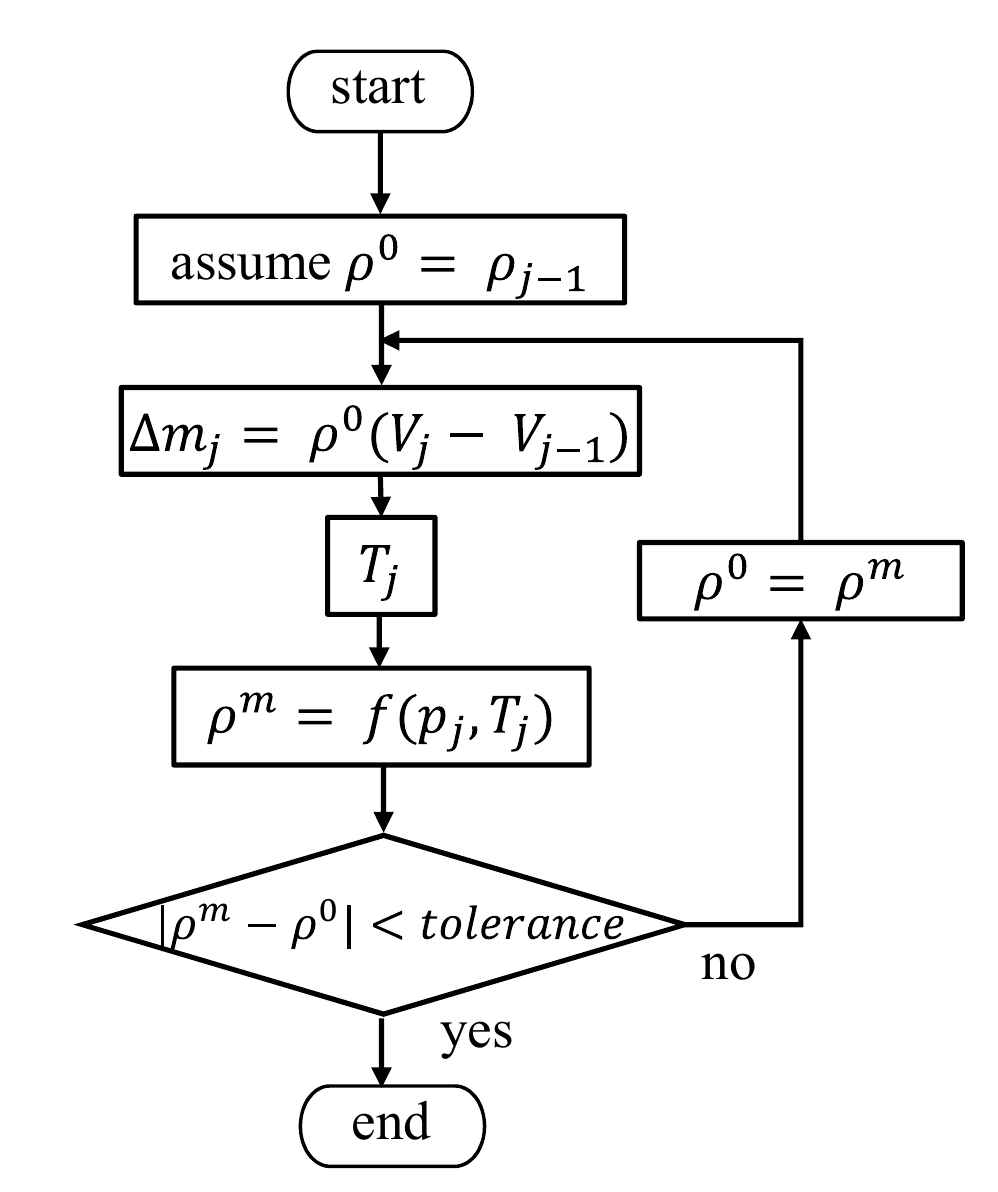}
  \caption{Flowchart of the iteration calculation on the mass flow rate.}
  \label{fig:delM}
\end{figure}

The equation \ref{eq:energy} and \ref{eq:mass} are the two differential equations that solve the thermodynamic process of gas inside the suction chamber. The mass entering or leaving the control volume can be derived from the pressure measured in the suction chamber. As shown in figure \ref{fig:delM}, the initial condition of the gas density $\rho^0$ is assumed to be equal to the gas density at the previous step $\rho_{j-1}$. The mass flow rate can be calculated as $\Delta m_{j} = \rho^{0} (V_{j} - V_{j-1})$, and the gas density $\rho^m$ at the $j$-th step can be derived by using equation \ref{eq:energy}. If the absolute difference between $\rho^0$ and $\rho^m$ is larger than a tolerance, $\rho^m$ is used as the initial condition for the next iteration until the difference is less than the tolerance.

The loss efficiency $\lambda_{sh}$ due to suction gas heating can be expressed as
\begin{equation}
    \lambda_{sh} = (\rho_{s} - \rho_{sh}) / \rho_s \,,
\end{equation}
where $\rho_{sh}$ is the gas density considering only the suction gas heating, $\rho_{s}$ is the density at the suction reference condition.

\subsubsection{Leakage loss}

In a properly designed compressor, the gaps between the roller face and the inner cylinder wall, and in the vane slot path are filled with lubricant, which acts as a sealant. As a result, the main leakage paths can be categorized as leakages past the radial clearance between the roller and the cylinder, as well as leakages past the vane tip. 

The leakage passage through the radial clearance between the roller and the cylinder can be modeled as a converging-diverging nozzle. Assuming that the leakage flow is only driven by the pressure difference, the average leakage flow rate $\dot{m}_{Lrc}$ through the radial clearance can be predicted as\citep{pandeya1978rolling}:

\begin{equation}
\dot{m}_{Lrc}=\frac{N_{rps} H}{\omega} \int_{\alpha_s}^{2 \pi-\alpha_d} \delta_{rc}(\alpha) p_{dt}(\alpha) \sqrt{\frac{\gamma}{R T_{dt}(\alpha)}\left(\frac{2}{\gamma+1}\right)^{\frac{(\gamma+1)}{(\gamma-1)}}} d \alpha \,,
\end{equation}
where $N_{rps}$ is the rotational speed in Hz, $\delta_{rc}$ is the minimum radial clearance between the roller and the cylinder, $H$ is the roller height, $p_{dt}$ and $T_{dt}$ are the pressure and temperature in the compression chamber, respectively, and $\gamma$ is the adiabatic index changed with pressure and temperature in general.

In this study, $\gamma$ under the isentropic condition can be estimated as\cite{wakabayashi1982analysis}

\begin{equation}
\gamma=\frac{\ln{\left(p_d / p_s\right)}}{\ln{\left(v_d / v_s\right)}} \,,
\end{equation}
where $p_d$ is the reference discharge pressure, $p_s$ is the reference suction pressure, $v_s$ is the specific volume at the reference suction condition, and $v_d$ is the specific volume at the reference discharge condition.  

Similarly, the leakage flow rate through the vane edges $\dot{m}_{Lve}$ can be described as

\begin{equation}
\dot{m}_{Lve}=\frac{N_{rps} }{\omega} \int_{\alpha_s}^{2 \pi} \delta_{vc}(\alpha) H_v(\alpha) p_{dt}(\alpha) \sqrt{\frac{\gamma}{R T_{dt}(\alpha)}\left(\frac{2}{\gamma+1}\right)^{\frac{(\gamma+1)}{(\gamma-1)}}} d \alpha \,,
\end{equation}
where $\delta_{vc}$ is the clearance between the vane edge and the cylinder head, which is often the same as the clearance between the roller faces and the cylinder head, and $H_v$ the vane extension inside the cylinder.

The total leakage mass flow rate can be calculated as 

\begin{equation}
\dot{m}_{Llk} = \dot{m}_{Lrc} + \dot{m}_{Lve}
\end{equation}

The loss in volumetric efficiency due to the leakage $\lambda_{lk}$ can be expressed as

\begin{equation}
    \lambda_{lk} = \frac{\dot{m}_{Llk}}{\rho_{s}N_{rps}V_2} 
\end{equation}

\subsubsection{Clearance volume loss}

Instead of solving mass and thermodynamic equations, the mass flow loss $\dot{m}_{Lcv}$ due to the clearance volume can be calculated by using the following equation:

\begin{equation}
\dot{m}_{Lcv}= N_{rps}\left(\rho_d-\rho_{2}\right) V_{\text {clearance}} \,,
\end{equation}
where $V_{\text {clearance}}$ is the clearance volume, $\rho_2$ is the density inside the suction chamber after the suction process and $\rho_d$ is the density of the discharged gas.

The loss in volumetric efficiency due to the clearance volume $\lambda_{cv}$ can be written as

\begin{equation}
    \lambda_{cv} = \frac{\dot{m}_{Lcv}}{\rho_{s}N_{rps}V_2}  
\end{equation}

\subsubsection{Back flow loss}

The delayed closing of the valves at the end of the discharge process may cause the discharged gas of high pressure back into the clearance volumes and into the suction chamber, by which the compressor volumetric efficiency decreases. When the throttling effects are considered the mass flow rate of the backflow $\dot{m}_{Lbf}$ through the gaps of the valve and the exhaust port can be calculated as \cite{liu1993simulation,lenz1994application}

\begin{equation}\label{eq:mbf}
    \begin{split}
    &\dot{m}_{Lbf}= \frac{N_{rps} }{\omega}\int_{\alpha_o}^{\alpha_c} \epsilon_b(\alpha) A_b(\alpha) \sqrt{\frac{2 \gamma}{\gamma-1} \rho_d p_d\left[\left(\frac{p_d}{p_{st}(\alpha)}\right)^{\frac{2}{\gamma}}-\left(\frac{p_d}{p_{st}(\alpha)}\right)^{\frac{\gamma+1}{\gamma}}\right]} d \alpha\\
    \end{split}
\end{equation}
where $\alpha_o$ and $\alpha_c$ are the open and close angle of the discharge valve, respectively, $\epsilon_b$ is the flow coefficient which is the result of flow through the valve and discharge port, $A_{b}$ is the effective flow area, and $p_{st}$ is the pressure inside the suction chamber.

If $ \quad \frac{p_d}{p_{st}(\alpha)} < \left(\frac{2}{\gamma+1}\right)^{\frac{\gamma}{\gamma-1}}$, $\dot{m}_{Lbf}$ under the critical condition can be expressed as
\begin{equation}\label{eq:mbf_lj}
   \dot{m}_{Lbf} = \frac{N_{rps} }{\omega}\int_{\alpha_o}^{\alpha_c} \epsilon_b(\alpha) A_b(\alpha) \left(\frac{2}{\gamma+1}\right)^{\frac{1}{\gamma-1}} \sqrt{\frac{2 \gamma}{\gamma+1} \rho_d p_d} d \alpha\\
\end{equation}

The loss in volumetric efficiency due to the backflow $\lambda_{bf}$ can be written as

\begin{equation}
    \lambda_{bf} = \frac{\dot{m}_{Lbf}}{\rho_{s}N_{rps}V_2} 
\end{equation}

\subsubsection{Mass flow gain due to supercharging}
Pressure fluctuations are inherent to the compressor due to the intermittent flow through the suction and discharge passages. The supercharging effect occurs when the pressure in the suction port is higher than that in the suction pipe inlet. When the suction chamber is supercharged, an additional amount of gas is delivered into the cylinder, causing the actual mass flow rate to exceed the ideal mass flow rate. The mass gain resulting from the supercharging effect can be expressed as
\begin{equation}
    \dot{m}_{sc} = N_{rps} (\rho_{sc} - \rho_{s}) V_{2} \,,
\end{equation}
where $\dot{m}_{sc}$ is the mass flow gain due to the supercharging effect, $\rho_{sc}$ is the refrigerant density considering only the supercharging effect, $\rho_{s}$ is the density at the suction reference condition.

The loss in the volumetric efficiency $\lambda_{sc}$ due to the supercharging effect can be expressed as
\begin{equation}
    \lambda_{sc} = (\rho_{s} - \rho_{sc} )/\rho_s \,,
\end{equation}
where the negative valve of $\lambda_{sc}$ means increases in the volumetric efficiency.

\subsection{Power losses during the cylinder process}\label{sec:Ploss}

\begin{figure}[h!]
  \centering
  \includegraphics[width=0.55\textwidth]{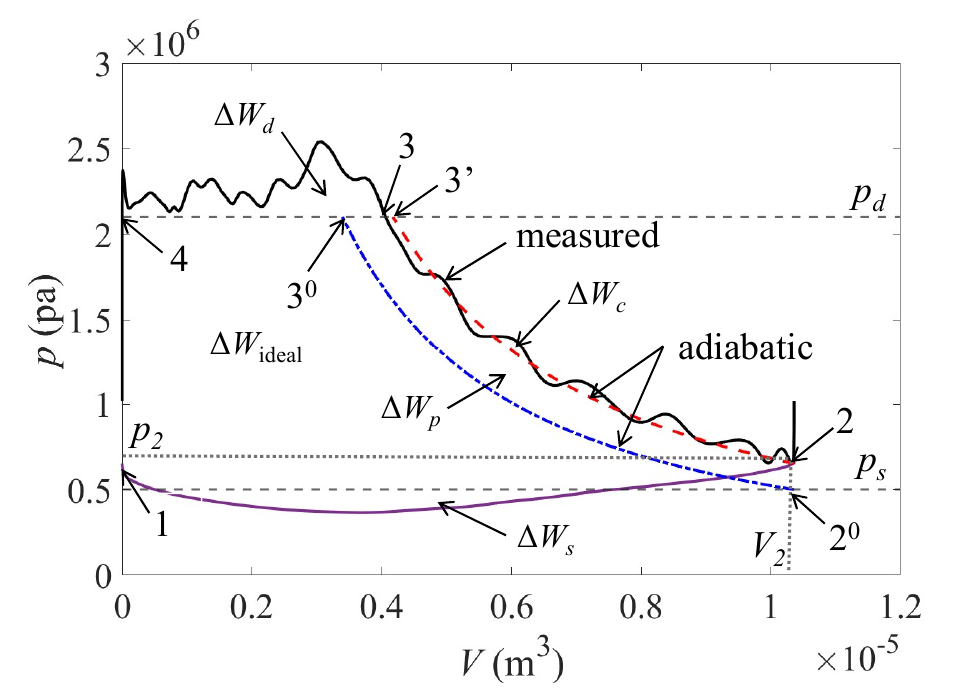}
  \caption{Example of P-V diagram.}
  \label{fig:PVexm}
\end{figure}

Power losses during the cylinder process can be quantified by comparing the ideal P-V diagram with the rea; P-V diagram. Figure \ref{fig:PVexm} shows a typical example of a P-V diagram. The real P-V diagram can be measured through the experiment, and power losses can be indicated by the areas enclosed by curves between the real and ideal processes. The power losses can be classified into four categories: power loss during the suction process $\Delta W_{s}$, power loss during the discharge process $\Delta W_{d}$, power loss during the compression process $\Delta W_{c}$, and power loss due to supercharging effect $\Delta W_{p}$. Therefore, the real work required for the compressor can be expressed as the ideal work plus all the power losses\cite{liu1993simulation}:

\begin{equation}
    W_{\text{real}} = W_{\text{ideal}} + \Delta W_{s} + \Delta W_{d} + \Delta W_{c} + \Delta W_{p}
\end{equation}

The ideal process for the power consumption can be calculated as

\begin{equation}
W_{\text {ideal }}=N_{rps}p_s V_{2}\left(\frac{\gamma}{\gamma-1}\right)\left[\left(\frac{p_d}{p_s}\right)^{\frac{\gamma-1}{\gamma}}-1\right]
\end{equation}

\subsubsection{Power loss in the suction process}

The power loss in the suction process is caused by the suction pressure that is lower than the reference suction pressure $p_s$, which can be expressed as

\begin{equation}
\Delta W_{s}= N_{rps} \int_{\alpha_s}^{2 \pi+\alpha_s}\left(p_{st}(\alpha)-p_s\right) \frac{d V}{d \alpha} d \alpha
\end{equation}

\subsubsection{Power loss in the discharge process}

The power loss during the discharge process can be quantified by the discharge pressure that is higher than the reference discharge pressure $p_d$, which can be expressed as the following equation: 

\begin{equation}
\Delta W_{d}= N_{rps} \mathop{\int}\limits_{discharge}\left(p_{d}-p_{dt}(\alpha)\right) \frac{d V}{d \alpha} d \alpha
\end{equation}

The power loss during the discharge process can be classified into discharge valve loss and discharge gas pulsation loss. Discharge valve loss refers to the additional work required to push the gas out of the cylinder due to the dynamic effect of the discharge valve. The discharge gas pulsation loss, on the other hand, is caused by the oscillation of the discharge pressure, which causes the back pressure temporarily higher than the reference discharge pressure. 
 
\subsubsection{Power loss in the compression process}

The power loss during the compression process can be considered as the difference between the real process and the adiabatic process, which can be written as 

\begin{equation}
    \begin{aligned}
    \Delta W_c = N_{rps} \Bigg( W_{\text{real}} - (\Delta W_{s} + \Delta W_{d}) - \Bigg(p_2 V_{2}\left(\frac{\gamma}{\gamma-1}\right)\left[\left(\frac{p_d}{p_s}\right)^{\frac{\gamma-1}{\gamma}}-1\right]& \\
    +(p_2 - p_s) V_2 \Bigg) \Bigg)& 
    \end{aligned}
\end{equation}

\subsubsection{Effect of supercharging on the power consumption}

In the ideal compression process, the cylinder gas is compressed from the reference suction pressure $p_s$ to the reference discharge pressure $p_d$. However, in a realistic process, the compression work is influenced by supercharging effects, resulting in a greater amount of work required. This extra power consumption is necessary to compress the additional amount of gas that is pumped into the suction chamber due to the supercharging effects. Then, the extra power consumption can be expressed as

\begin{equation}
\Delta W_{p}= N_{rps} \left( \frac{\gamma p_{2} V_{2}}{\gamma-1}\left[\left(\frac{p_d}{p_{2}}\right)^{\frac{\gamma-1}{\gamma}}-1\right]-\frac{\gamma p_s V_{2}}{\gamma-1}\left[\left(\frac{p_d}{p_s}\right)^{\frac{\gamma-1}{\gamma}}-1\right]+\left(p_2-p_s\right) V_{2} \right)
\end{equation}

\section{Results} \label{sec:results}

\subsection{Overall performance}

\begin{figure}[h!]
  \centering
  \includegraphics[width=0.49\textwidth]{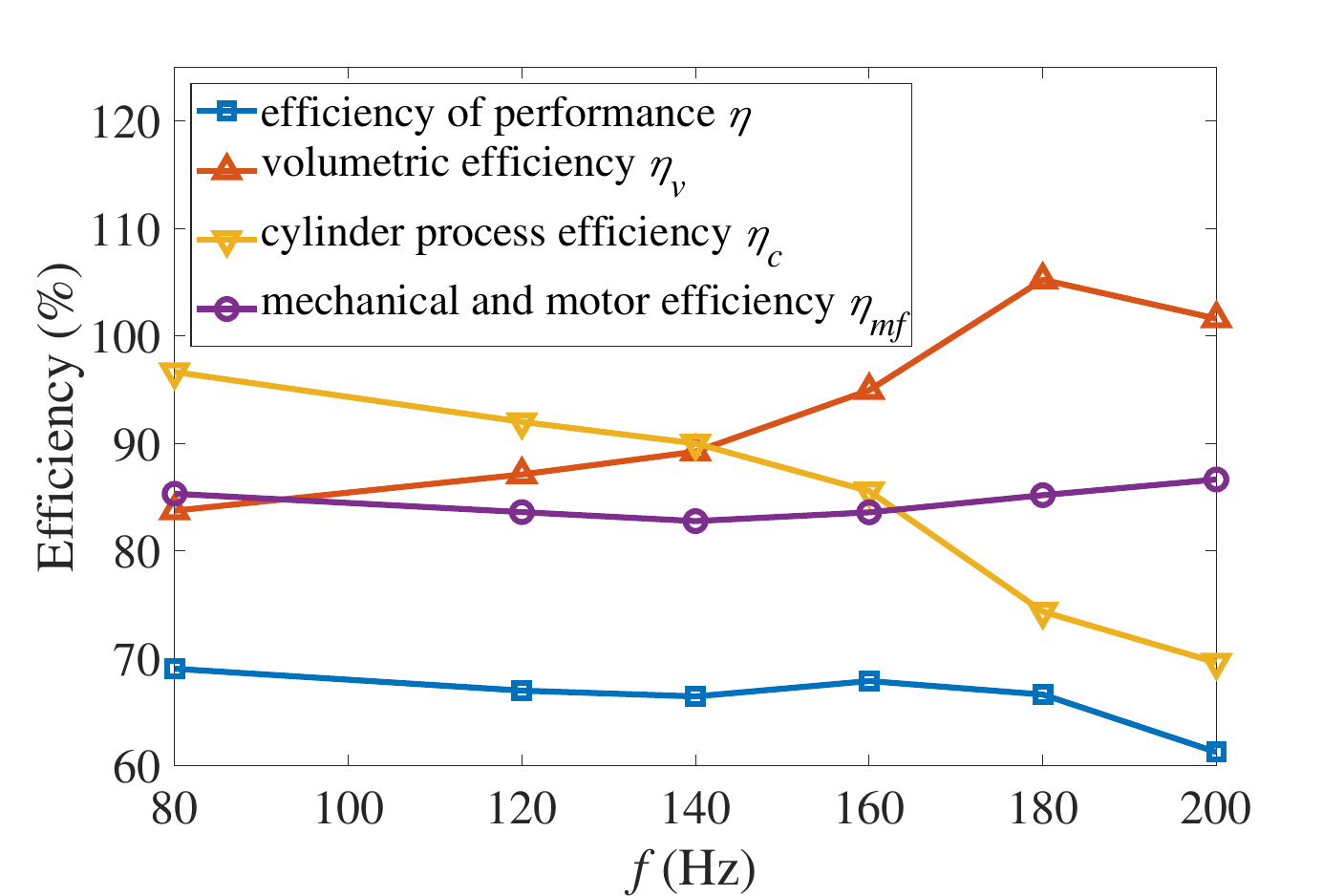}
  \caption{Comparison of performance efficiencies against rotational speeds.}
  \label{fig:Oeff}
\end{figure}

Figure \ref{fig:Oeff} shows the overall performance of the compressor at variable speeds. The overall performance efficiency exhibits a slight decrease from 80Hz to 140Hz. After a local peak of around 68\% at 160Hz, it shows a downward trend. Notably, there is a significant reduction in performance efficiency at 200Hz. 

In contrast, the volume efficiency continuously rises from approximately 84\% at 80Hz to 105\% at 180Hz, before dropping to about 101\% at 200Hz. The cylinder process efficiency experiences a continuous decline, with a significant drop occurring between 140Hz and 180Hz, corresponding to a substantial increase in volumetric efficiency in this speed range. This is reasonable, as more power is needed to compress the additional refrigeration gas. 

The mechanical and motor efficiency reaches a minimum of about 83\% at 140Hz, After which it steadily increases to around 87\% at 200Hz. This suggests that the motor efficiency may improve at a speed above 140Hz, indicating that high-speed operation (above 140Hz) may benefit mechanical and motor efficiency in a well-designed compressor. 

\subsection{Analysis of the volumetric efficiency loss}

\begin{figure}[h!]
  \centering
  \begin{subfigure}[b]{0.49\textwidth}
    \includegraphics[width=\textwidth]{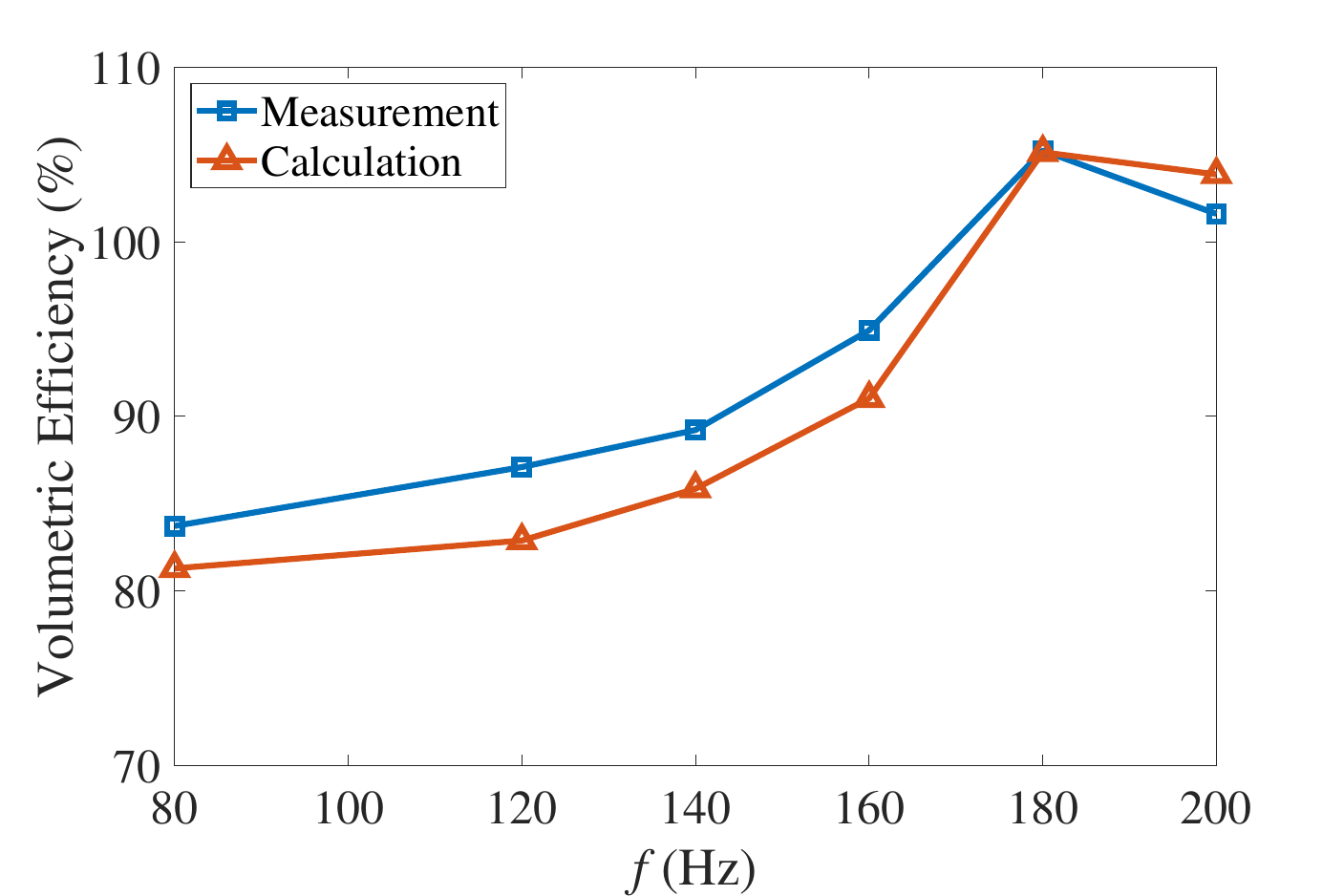}
    \caption{}
  \end{subfigure}             
  \begin{subfigure}[b]{0.49\textwidth}
    \includegraphics[width=\textwidth]{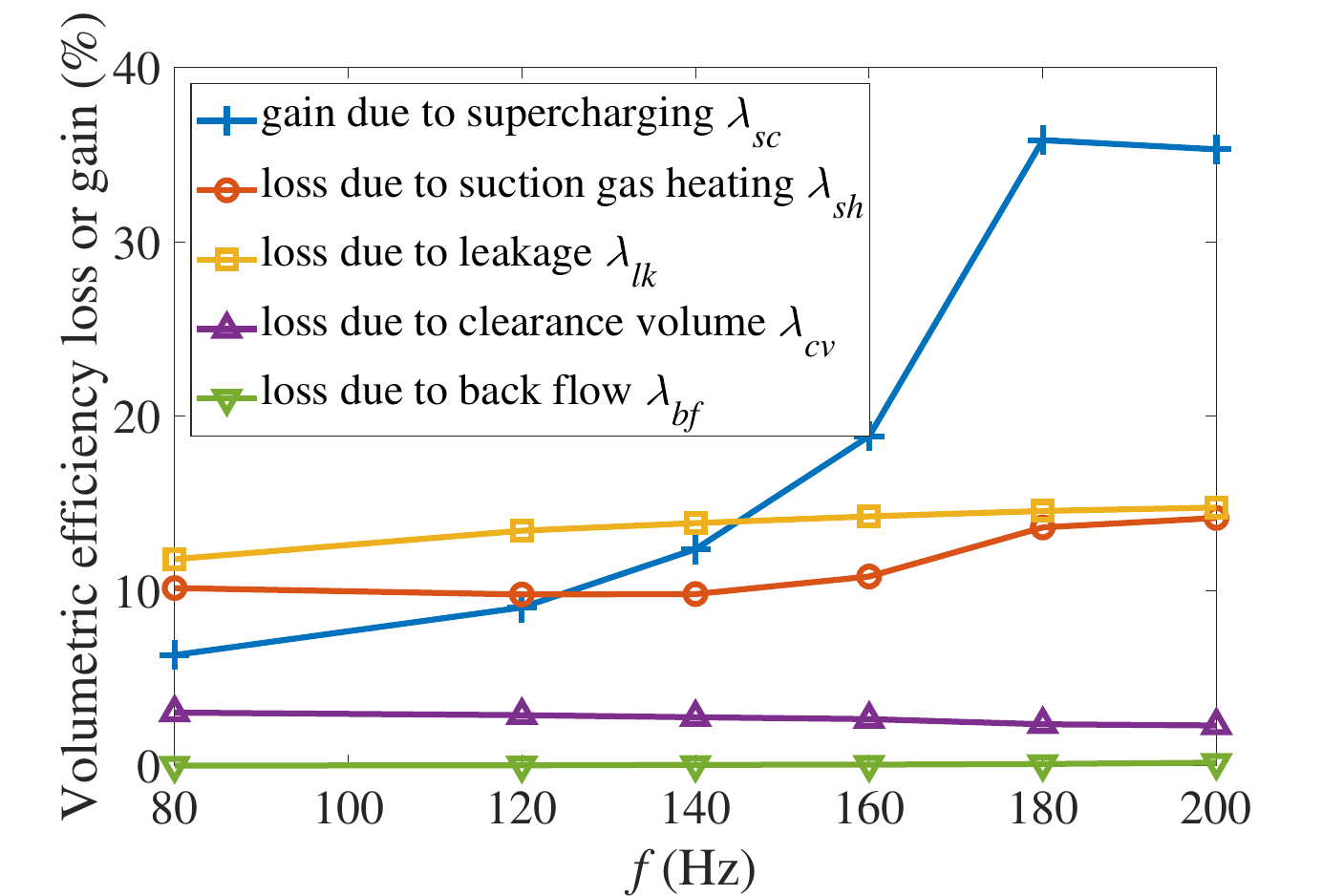}
    \caption{}
  \end{subfigure}             
  \caption{The results of (a) comparison of volumetric efficiency between the measurement results and the calculation and (b) the volumetric efficiency loss for each factor.}
  \label{fig:VEloss}
\end{figure}

Figure \ref{fig:VEloss} (a) compares the volumetric efficiency of the compressor between the measurement results and the results calculated using equation \ref{eq:VEcal}. The two loss curves show good agreement, with an average difference of less than 5\%, showing that the current loss modeling approach can accurately predict volumetric efficiency loss at different speeds. Figure \ref{fig:VEloss} (b) shows the volumetric efficiency loss due to different factors at various speeds. Based on the experimental data, each volumetric efficiency loss is calculated according to the mathematical modeling in Section \ref{sec:massLoss}.   

\begin{figure}[h!]
  \centering
  \begin{subfigure}[b]{0.49\textwidth}
    \includegraphics[width=\textwidth]{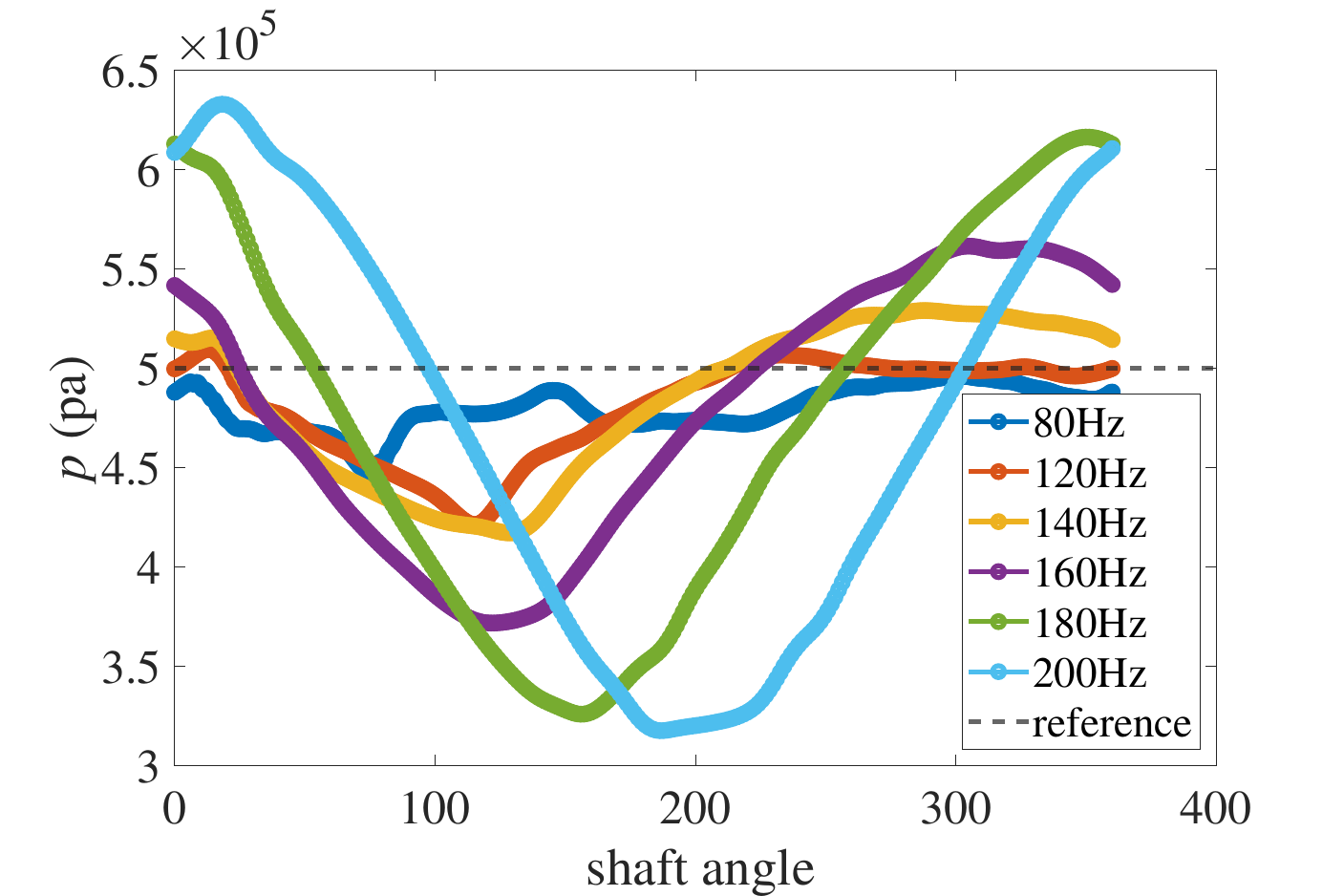}
    \caption{}
  \end{subfigure}             
  \begin{subfigure}[b]{0.49\textwidth}
    \includegraphics[width=\textwidth]{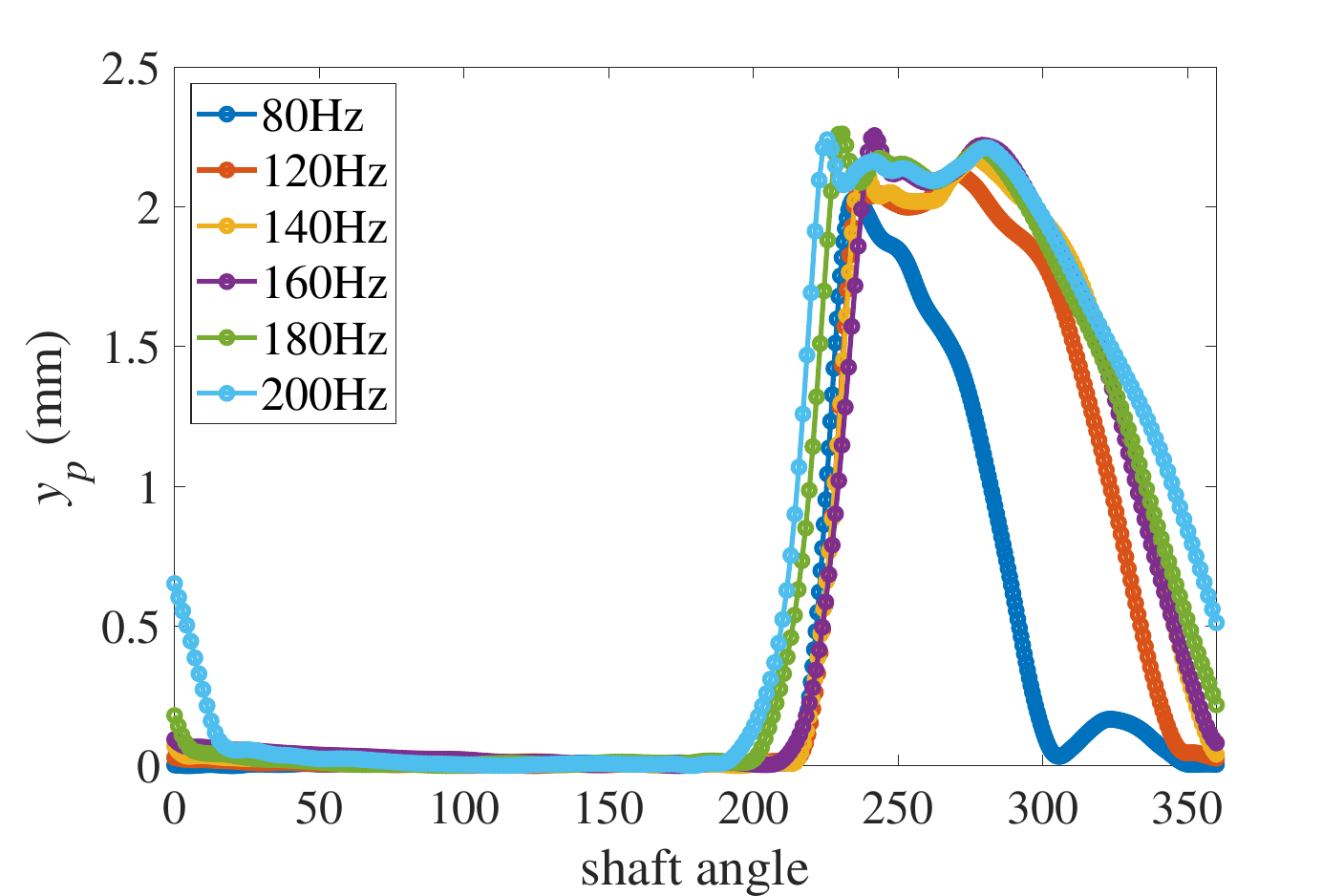}
    \caption{}
  \end{subfigure}             
  \caption{The measurement results of (a) suction pressure pulsation at variable speeds and (b) the discharge valve lift at variable speeds.}
  \label{fig:Spressure}
\end{figure}

As shown in figure \ref{fig:VEloss} (a), except at 200Hz $\lambda_{sc}$ has a significant and continuous increase, and the rate of increase becomes larger at higher speeds. The change in increase rate can be explained by examining the suction pressure as follows. Figure \ref{fig:Spressure} (a) presents the measurement results of the pressure change inside the suction chamber. All pressure curves follow a similar trend, with pressure dropping at the start of suction and then rising to a maximum value at the end of the process. The pressure fluctuation becomes more pronounced at higher speeds, which corresponds to a larger pressure at the end of the suction process. Although the suction pressure at 200Hz has the highest rate of increase at the end of the suction process, it does not reach its peak at that point. In contrast, the suction pressure at 180Hz reaches its peak close to the end of the suction process, causing its value larger than that at 200Hz. As a result, $\lambda_{sc}$ and $\eta_{v}$ decrease slightly from 180Hz to 200Hz.

The results for $\lambda_{lk}$ and $\lambda_{sh}$ in figure \ref{fig:VEloss} (a) indicate that leakage and suction heating are primary contributors to volumetric efficiency loss. The leakage loss becomes more severe at higher speeds, reaching a maximum value of 15\% at 200Hz. The suction heating loss remains almost unchanged at speeds lower than 140Hz, but increases from 10\% to about 14\% at 200Hz. 

As the clearance volume is determined by the compressor's geometric structure, $\lambda_{cv}$ remains almost constant at 3\% across the measured speed range. In contrast, $\lambda_{bf}$ is almost negligible. Figure \ref{fig:Spressure} (b) illustrates the measured displacement of the discharge valve during the tests. At speeds below 200Hz, the valve closes promptly before 360$^{\circ}$. At 200Hz, the slight delay in valve closure, combined with the throttling effect caused by a small valve displacement, results in a very small amount of backflow. Consequently, there is no discharge of gas back into the suction chamber, and hence, no backflow loss. It is worth noting that the discharge valve experiences delayed opening at higher speeds, which is influenced by the faster compression process and the valve's inertia.

\subsection{Analysis of the power loss}

\begin{figure}[h!]
  \centering
  \begin{subfigure}[b]{0.49\textwidth}
    \includegraphics[width=\textwidth]{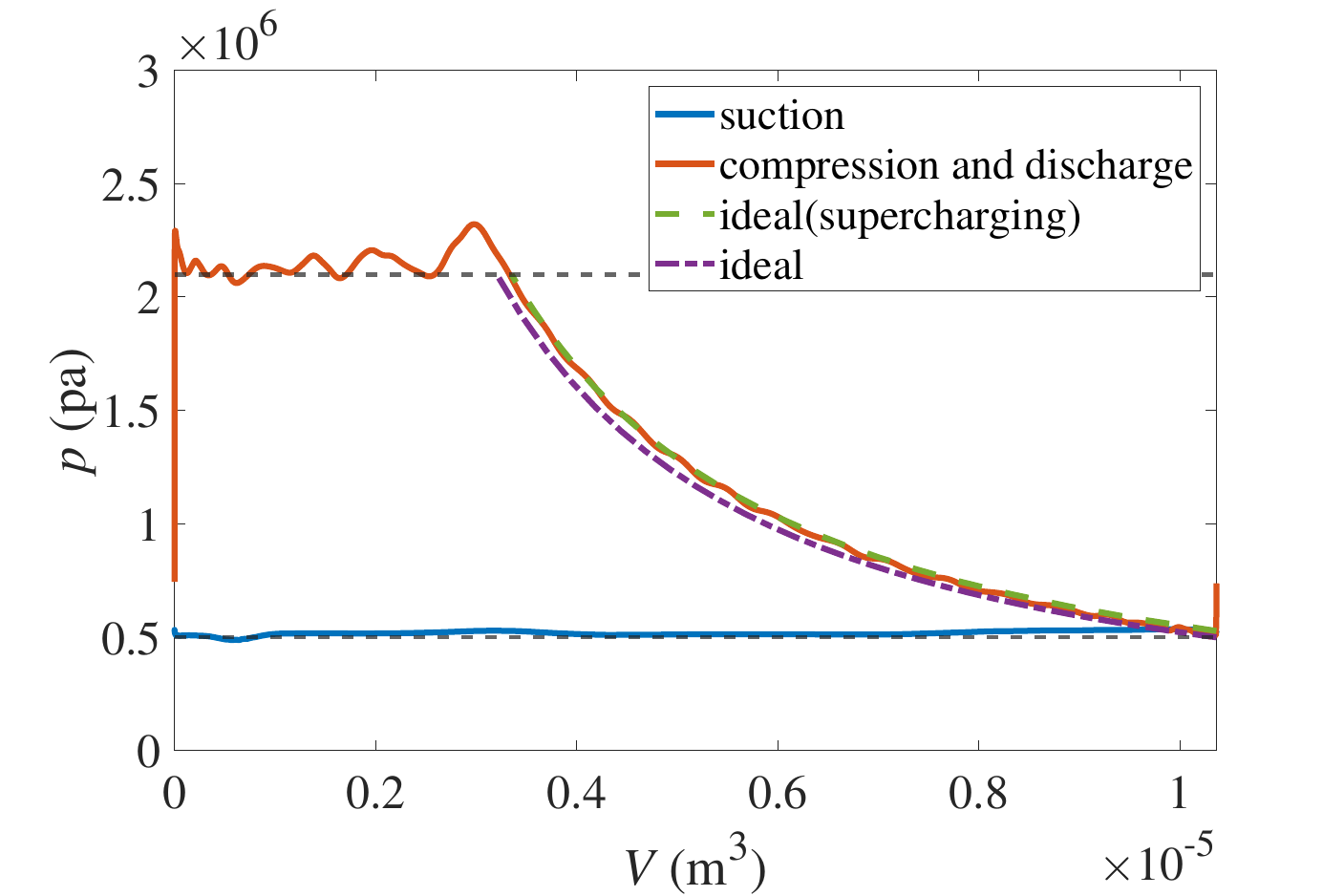}
    \caption{}
  \end{subfigure}
  \begin{subfigure}[b]{0.49\textwidth}
    \includegraphics[width=\textwidth]{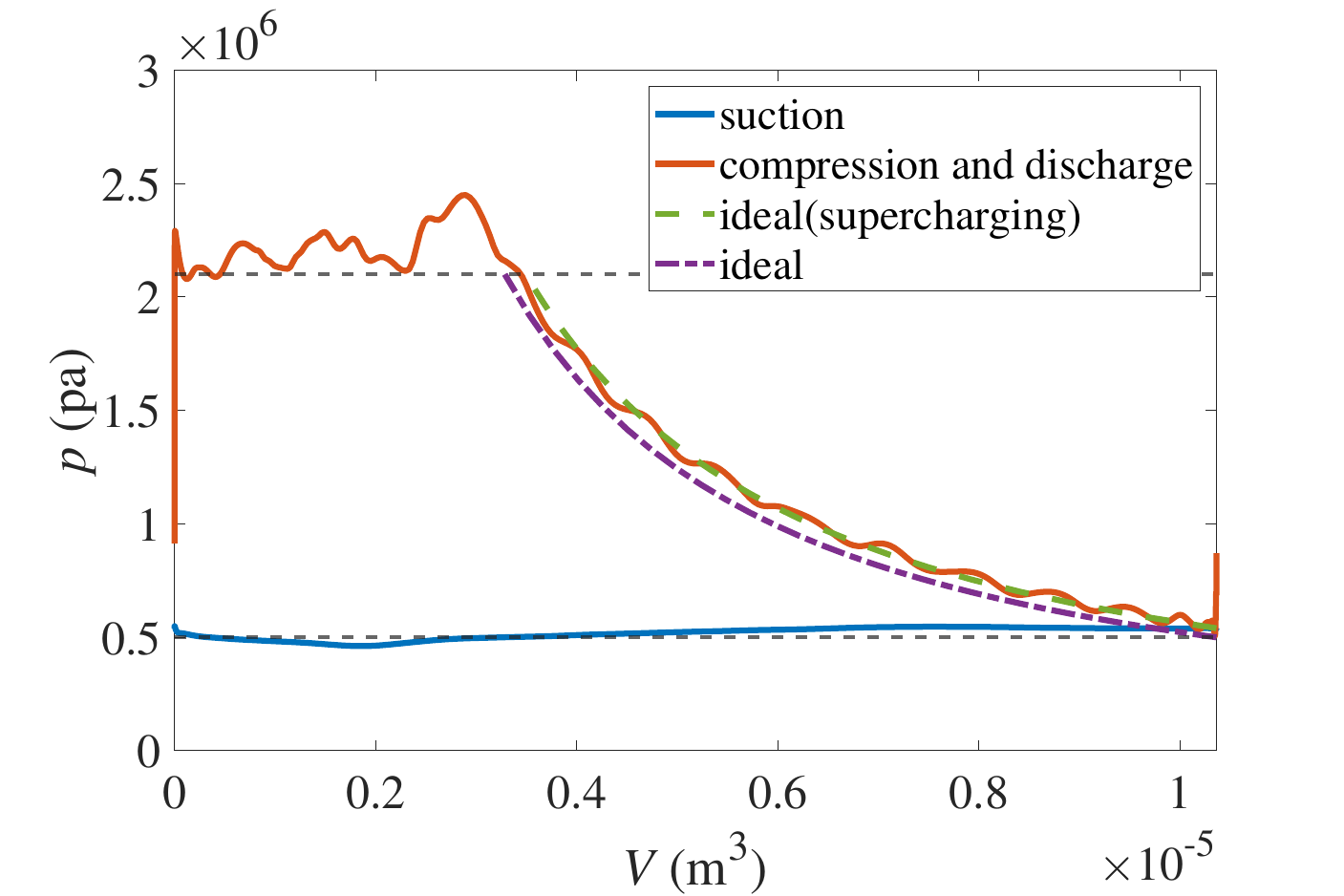}
    \caption{}
  \end{subfigure}             
  \begin{subfigure}[b]{0.49\textwidth}
    \includegraphics[width=\textwidth]{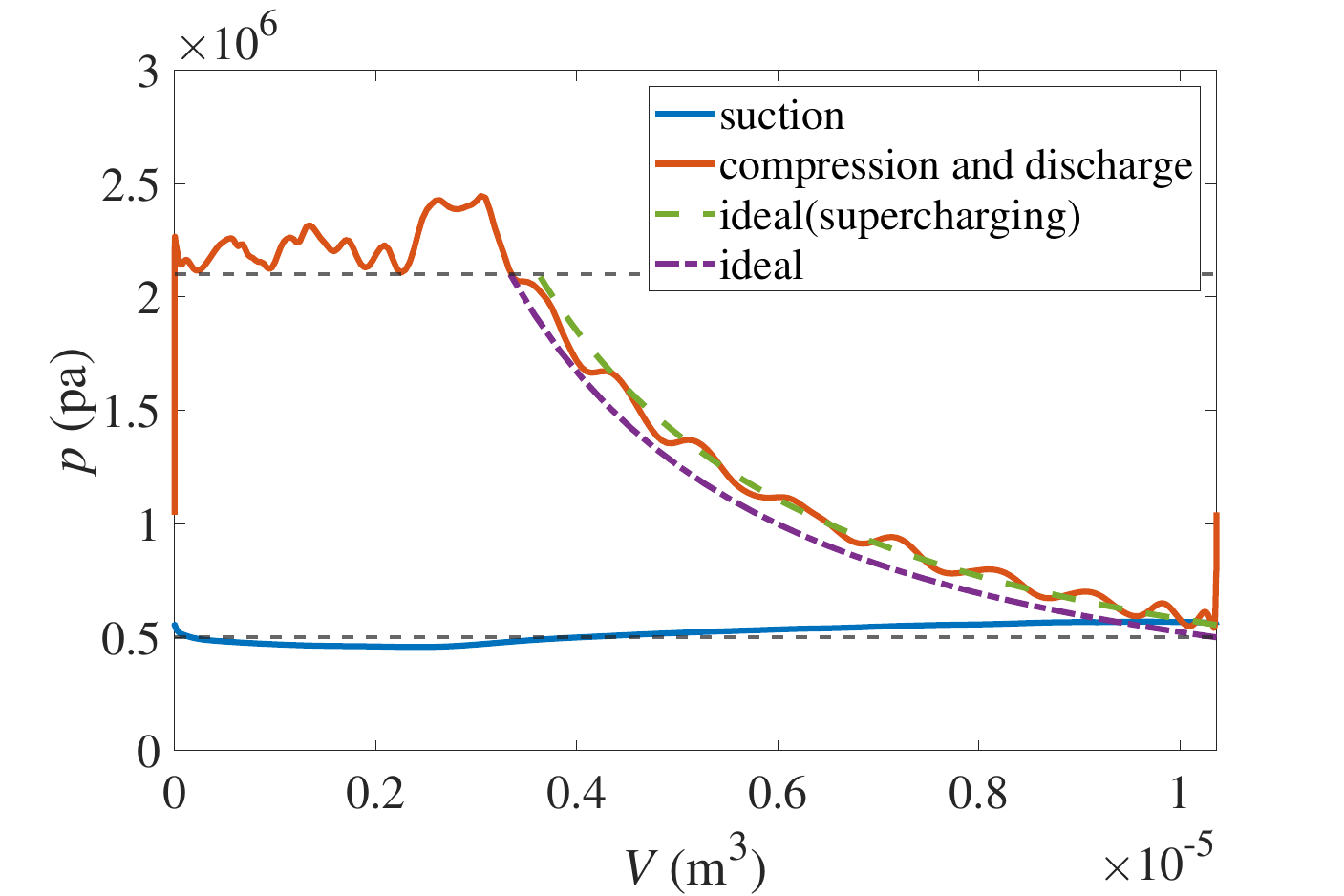}
    \caption{}
  \end{subfigure}             
  \begin{subfigure}[b]{0.49\textwidth}
    \includegraphics[width=\textwidth]{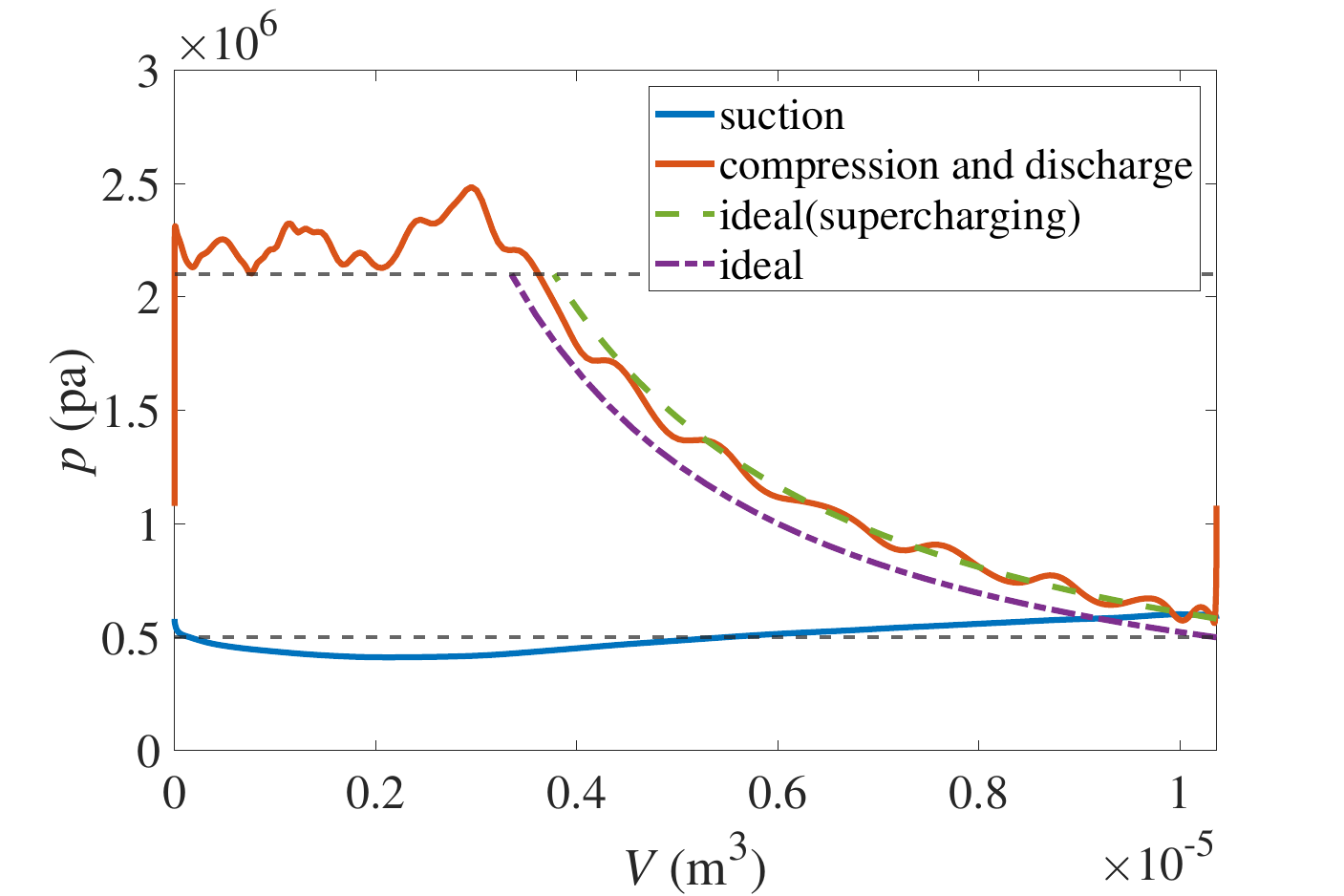}
    \caption{}
  \end{subfigure}             
  \begin{subfigure}[b]{0.49\textwidth}
    \includegraphics[width=\textwidth]{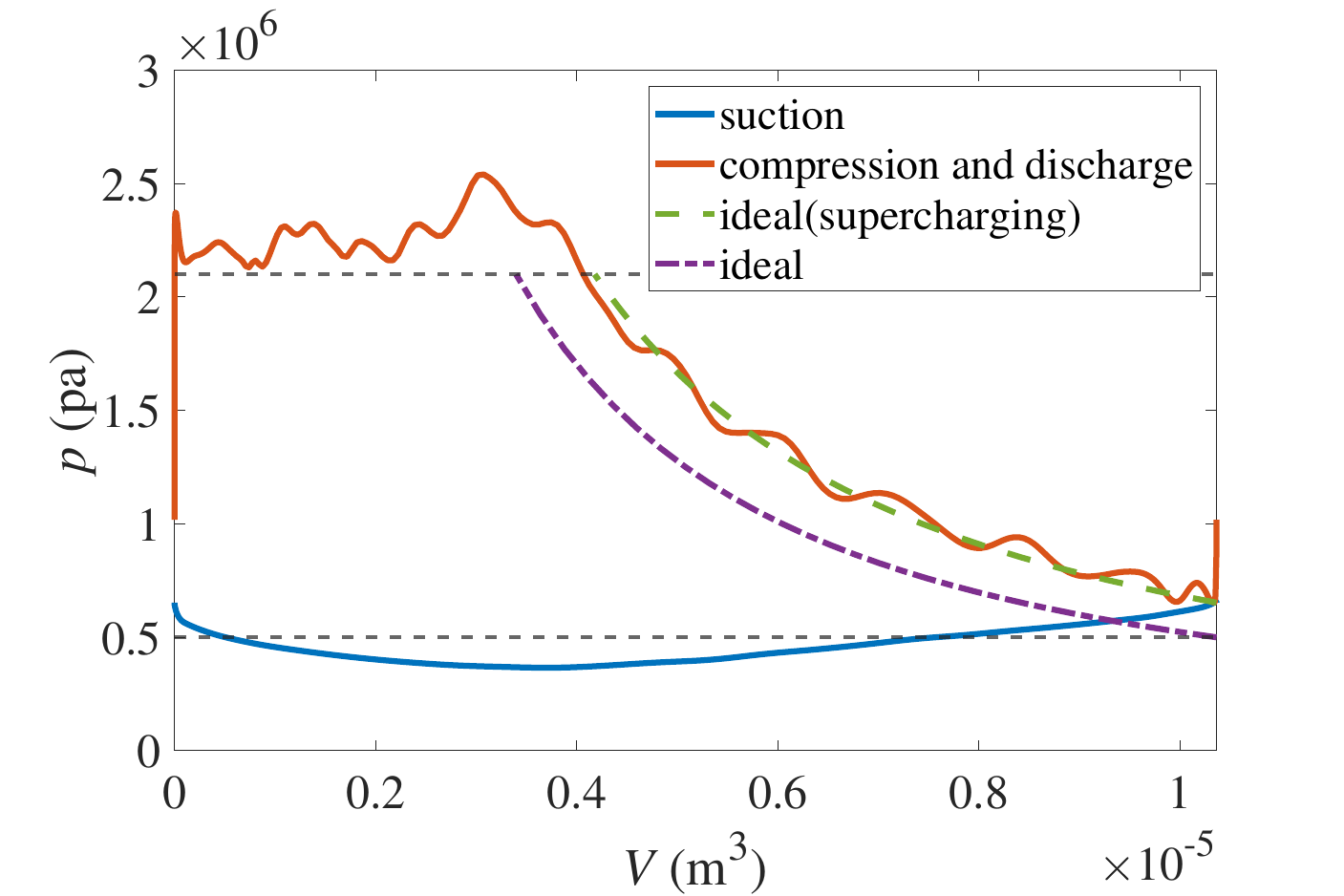}
    \caption{}
  \end{subfigure}             
  \begin{subfigure}[b]{0.49\textwidth}
    \includegraphics[width=\textwidth]{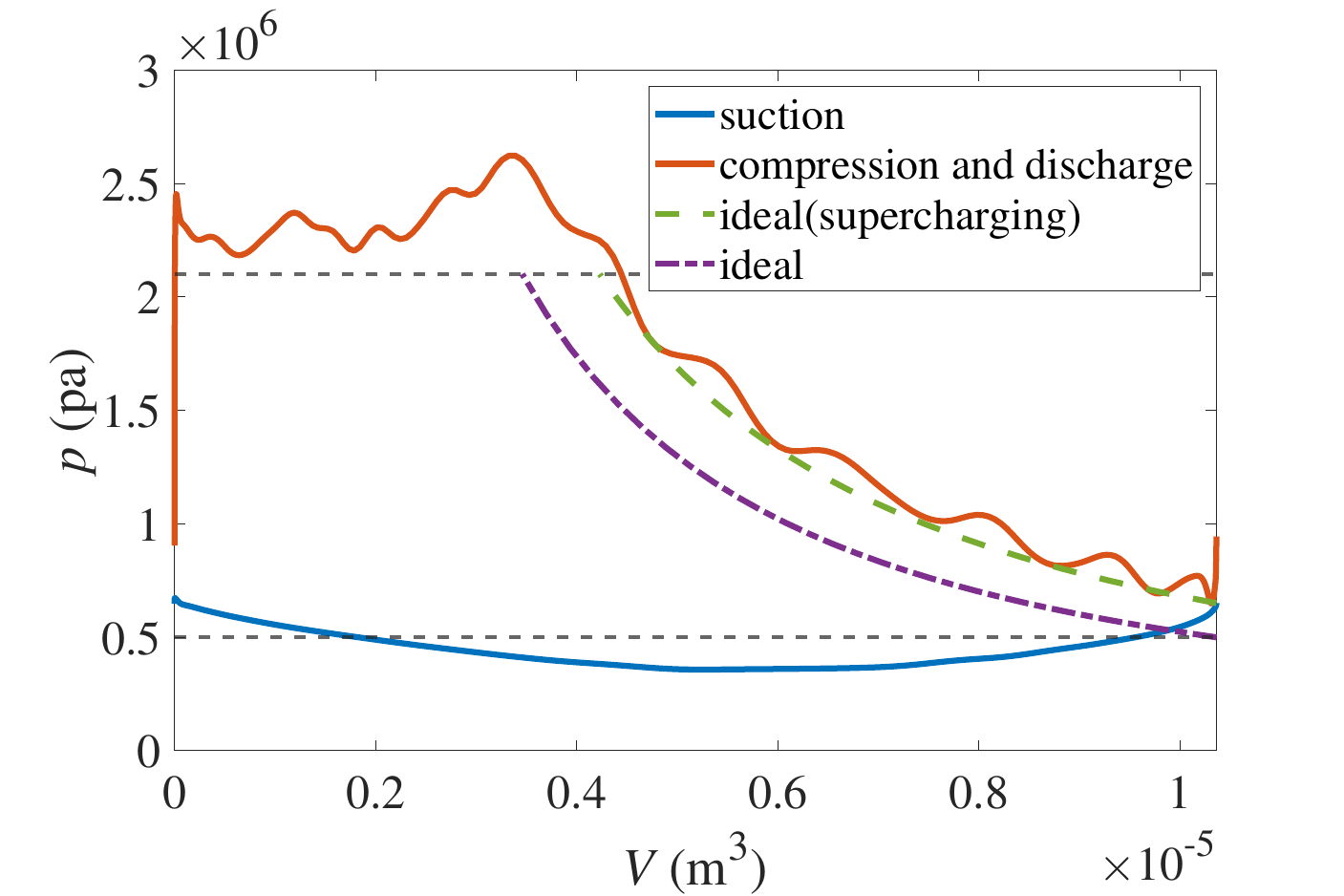}
    \caption{}
  \end{subfigure}             
  \caption{The P-V diagram at speeds of (a) 80Hz, (b) 120Hz, (c) 140Hz, (d) 160Hz, (e) 180Hz, and (f) 200Hz.  }
  \label{fig:PV}
\end{figure}

Figure \ref{fig:PV} shows the P-V diagram of the compressor operating at different speeds. The suction pressure curve is measured from the pressure transducer at the inlet port, whereas the discharge pressure curve is measured from the pressure transducer at the discharge port. Two horizontal dashed lines represent the reference suction pressure and reference discharge pressure. The ideal compression curve is calculated under the isentropic compression started from the reference suction pressure, and the area enclosed by it and two reference lines represent the ideal gas compression work. The curve of ideal compression under supercharging is plotted from the real pressure $p_2$ at the end of the suction process, which stands for the isentropic compression taking into account the supercharging effect. At 80Hz, the P-V curves closely resemble the ideal process, with relatively stable pressure during the suction and compression processes. However, as rotational speeds increase, pressure fluctuations in the suction, compression, and discharge processes become more pronounced, which can be attributed to faster roller motion. The resulting pressure fluctuation during the suction process leads to a notable supercharging effect, causing the area between two ideal compression curves to be larger. Moreover, at higher speeds, the area above the discharge reference line increases, resulting in larger power loss in the discharge process. 

\begin{figure}[ht!]
  \centering
  \includegraphics[width=0.55\textwidth]{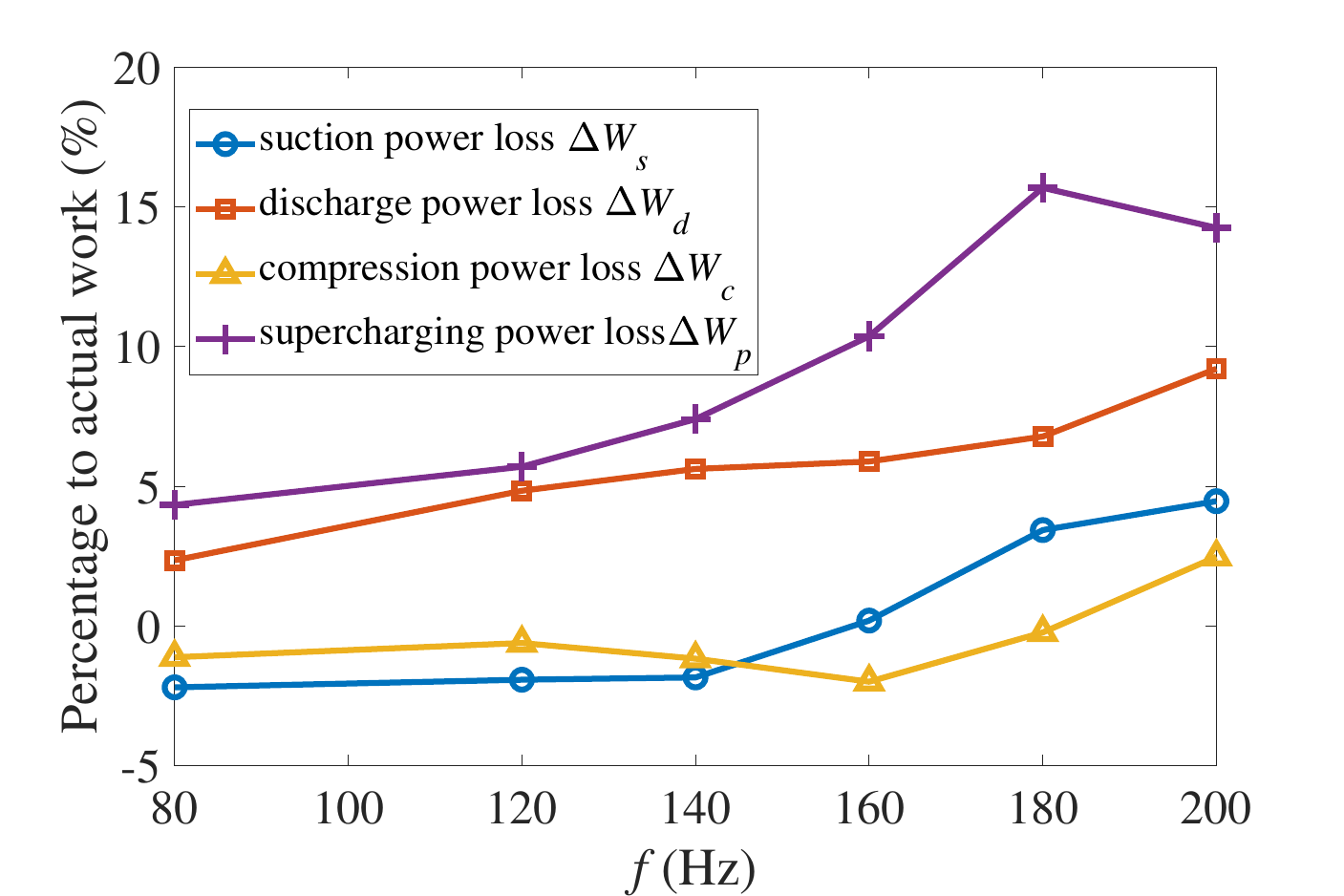}
  \caption{The results of power loss percentage of each factor.}
  \label{fig:Ploss}
\end{figure}

Figure \ref{fig:Ploss} summarizes the percentage ratio of each loss relative to the actual compression work. The extra work done under the supercharging effect is the most significant, and its value is proportional to the pressure at the end of the suction process. It is reasonable that extra compression works are consumed on the increased amount of gas under the supercharging effect. The second most important factor is the discharge loss. On one hand, the pressure in the suction chamber is always higher to overcome the flow resistance introduced by the discharge valve. On the other hand, an extra amount of work is done due to the presence of the discharge gas pulsation. At higher speeds, the fast discharge process implies that the resistance becomes higher due to the inertia of the discharge valve. Meanwhile, the discharge gas pulsation is enhanced owing to the larger flow rate at higher speeds. Therefore, the discharge loss increases at higher speeds.

The suction loss is negligible at speeds lower than 140Hz, but increases notably from 140Hz to 200Hz. At high speeds, the gas inertia prevents the gas from filling the suction volume fast enough during certain stages of the suction process, resulting in a suction pressure lower than the reference suction pressure. As a result, the compressor operating at high speeds experiences relatively large suction loss.
It is worth noting that the compression loss is negative at speeds below 180Hz. This may be due to the leakage from the compression chamber to the suction chamber, which reduces the amount of gas to be compressed. Another reason may be the heat transfer from the gas to the cylinder wall, as the gas temperature may exceed that of the cylinder wall at the late stage of the compression process. Despite the negative value, the compression loss still increases at 180Hz and 200Hz.

\newpage

\section{Discussions} \label{sec:discussions}

\subsection{Net effects of loss factors on the compressor performance}

To improve compressor performance, especially at high speeds, it is essential to identify the dominant factors contributing to efficiency loss. However, evaluating volumetric efficiency or power loss alone is insufficient to accurately assess the net effects of these loss factors. For example, the supercharging effect can improve volumetric efficiency but results in a noticeable power loss. Conversely, leakage can significantly decrease volumetric efficiency but simultaneously reduces power loss during the compression process. Therefore, a combined analysis of various loss factors is necessary to quantify their net effects accurately.

The net effects of loss factors on the compressor performance can be quantified as

\begin{equation}\label{eq:ratio}
\text{rate}_{\text{loss}} = \frac{\dot{m}_{\text {ideal }}+\dot{m}_{\text{loss}}}{W_\text{ideal }+\Delta W_{\text{loss}}} \,,
\end{equation}
where $\text{rate}_{\text{loss}}$ is the ratio of mass flow rate to power accounting for the net effect of each loss factor, $\dot{m}_{\text{loss}}$ is the mass rate loss caused by each loss factor and $\Delta W_{\text{loss}}$ is the power loss due to each loss factor.

\begin{figure}[h!]
  \centering
  \includegraphics[width=0.55\textwidth]{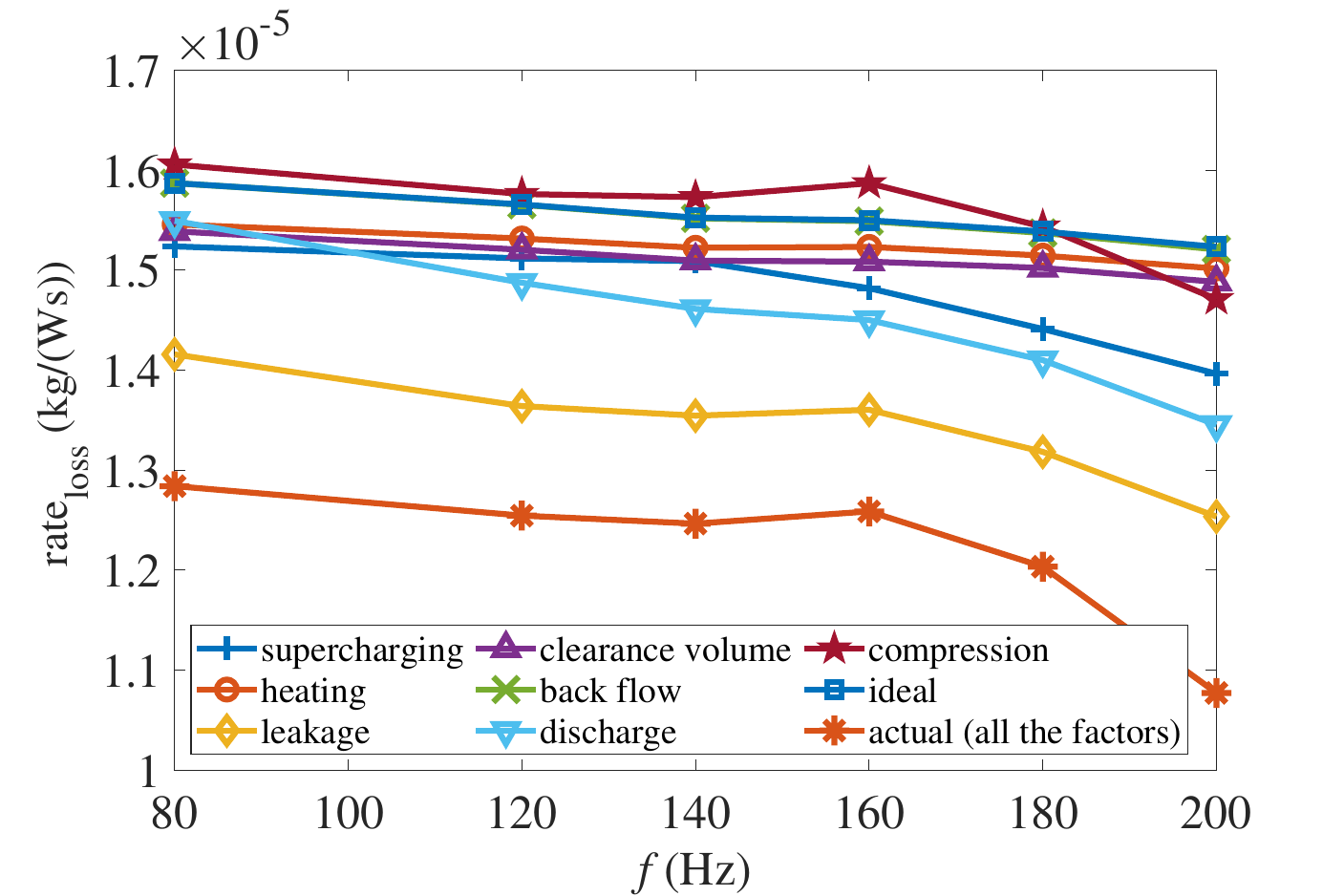}
  \caption{The net effects of loss factors at variable speeds.}
  \label{fig:ratio}
\end{figure}

Figure \ref{fig:ratio} shows the results of $\text{rate}_{\text{loss}}$ for each loss factor. The term `ideal' represents the compressor's performance under ideal operating conditions, without any loss factors. On the other hand, `actual' refers to the compressor's performance when all loss factors are taken into account. A value smaller than the ideal result represents the decrease in compressor performance caused by that particular loss factor. The smaller the value, the greater the contribution of that loss factor to the overall efficiency loss of the compressor.  

Within the measured speed range, the impact of leakage on compressor efficiency loss is the most significant. Due to the working principles of rotary compressors, there are various clearances that are determined by the design, manufacturing, and concentricity of the assembly \cite{yanagisawa1985leakage}. At speeds exceeding 160Hz, the energy efficiency degradation of the compressor caused by leakage becomes even more pronounced. Therefore, optimizing the internal dimensions of compressors to reduce leakage loss is suggested to greatly enhance compressor efficiency at high speeds. 

The second major factor contributing to compressor efficiency loss is discharge loss, which becomes more significant at high speeds. High-speed operation of the compressor increases the discharge mass flow rate\cite{monasry2018development}. However, due to factors such as the inertia and stiffness of the discharge valve\cite{prater2003optical}, there is a delay in the opening of the valve plate at high speeds (figure \ref{fig:Spressure} (b)), resulting in over-compression during the initial stages of the discharge process (figure \ref{fig:PV}). Additionally, discharge gas pulsation also contributes to discharge loss. To mitigate this, a high-speed discharge valve combined with a discharge system (such as a discharge port and discharge muffler) can be designed and optimized to reduce discharge loss at high speeds. The impact of suction heating and clearance volume on compressor performance is almost equal and independent of speeds. Suction heating loss is primarily caused by the heating of gas inside the suction pipe, which is influenced by temperature distributions, flow speed, and heat transfer coefficient. The clearance volume loss is determined by the compressor's structure. Considering their limited contributions to the compressor efficiency loss and the mechanisms of their generation, a limited effort is suggested can be made to reduce these two types of losses.   

\subsection{Net effects of supercharging on the compressor performance}

The supercharging effect caused by pressure pulsation during the suction process is the most significant characteristic of a compressor operating at high speeds. It significantly increases both the volumetric efficiency gain and power loss at high speeds. The net effect of the supercharging effect can be calculated using equation \ref{eq:ratio}, where $\dot{m}_{\text{loss}}$ is the mass flow gain and $\Delta W_{\text{loss}}$ consists of $\Delta W_{s}$ and $\Delta W_{p}$ due to the supercharging effect. The results shown in figure \ref{fig:ratio} demonstrate that the supercharging effect is detrimental to the compressor efficiency, which is consistent with the conclusion drawn by \citet{liu1993simulation}. At speeds above 140Hz, the impact of the supercharging effect on the compressor efficiency degradation grows to be increasingly significant, rendering it a third dominant factor. 

\begin{figure}[h!]
  \centering
  \begin{subfigure}[b]{0.49\textwidth}
    \includegraphics[width=\textwidth]{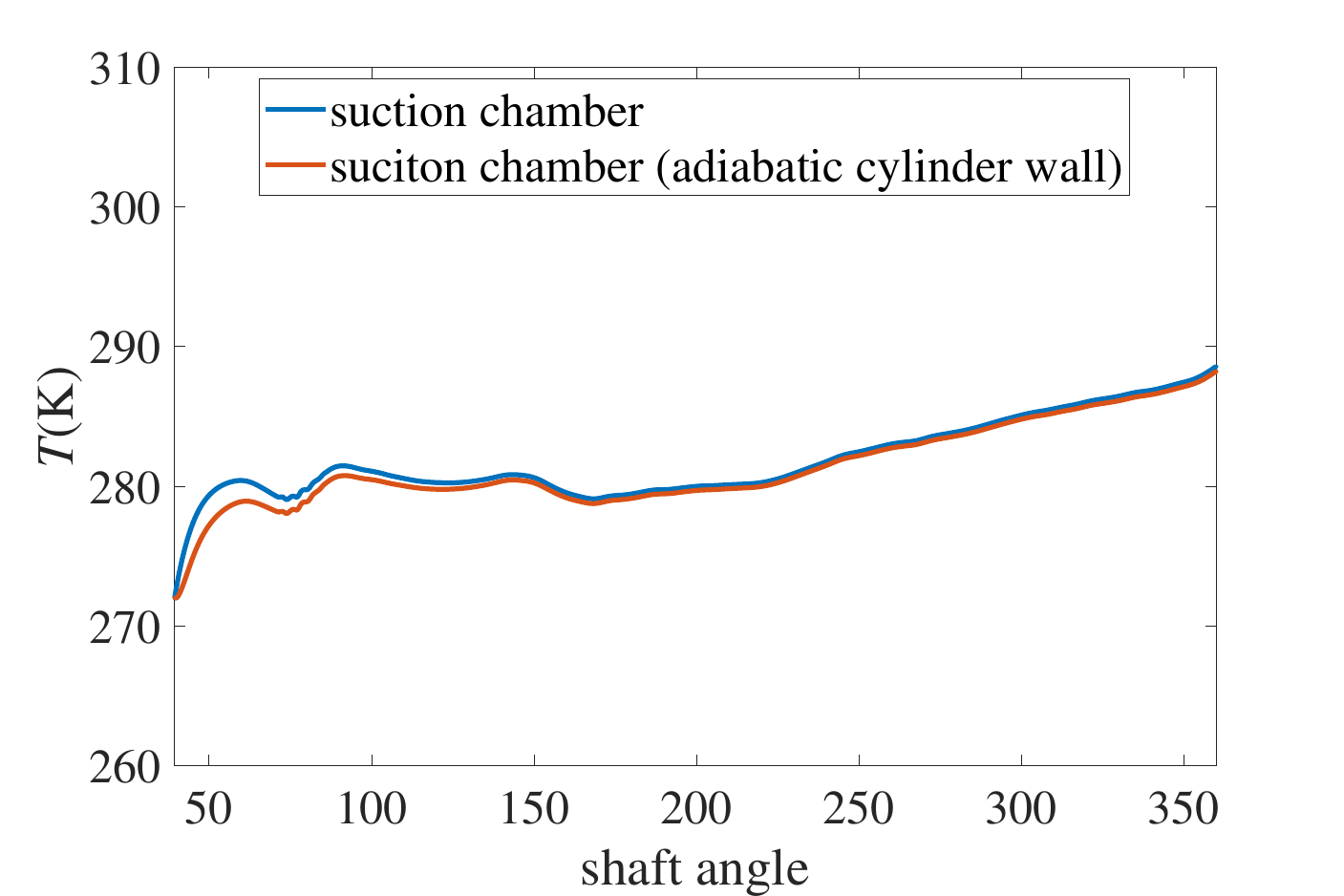}
    \caption{}
  \end{subfigure}
  \begin{subfigure}[b]{0.49\textwidth}
    \includegraphics[width=\textwidth]{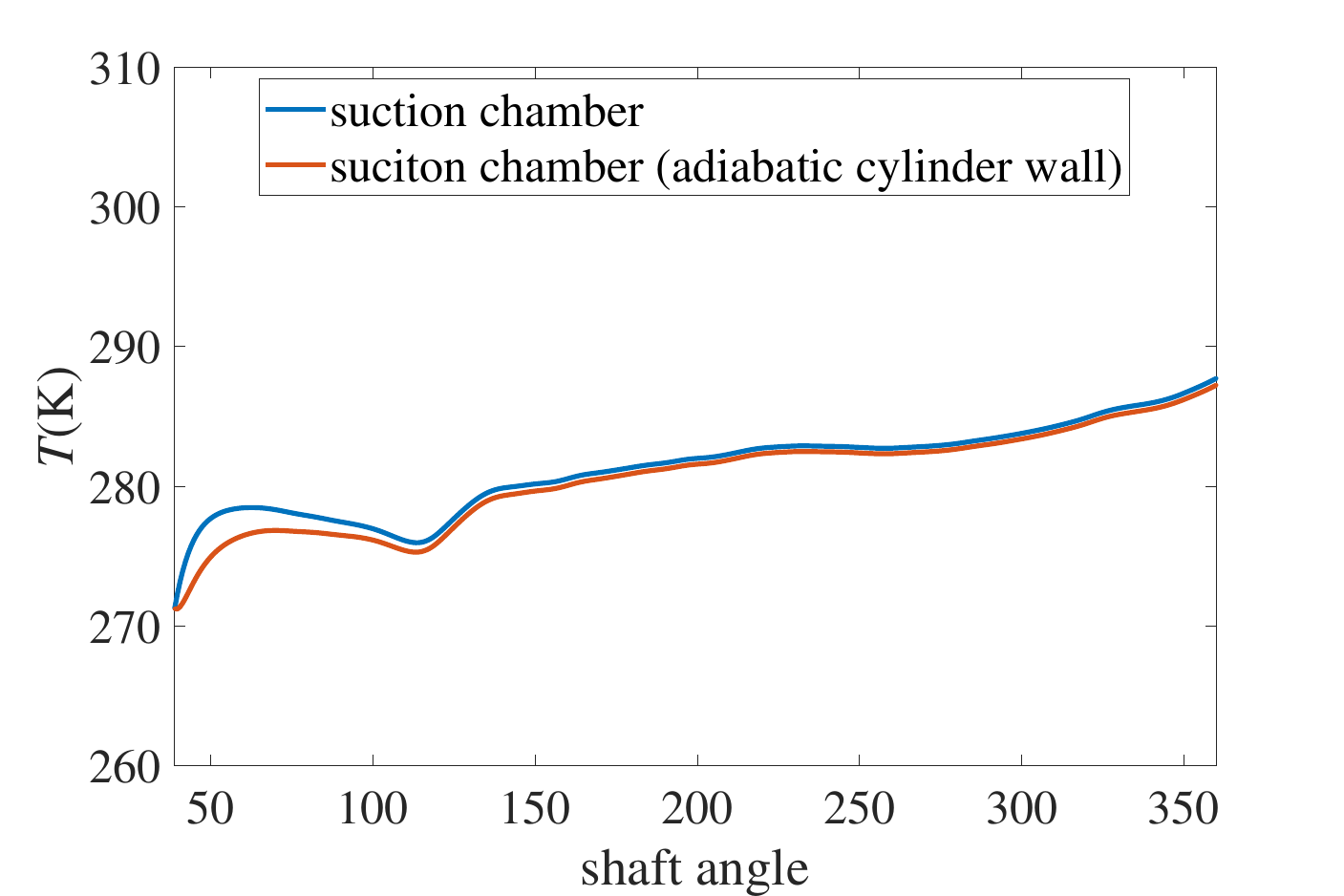}
    \caption{}
  \end{subfigure}             
  \begin{subfigure}[b]{0.49\textwidth}
    \includegraphics[width=\textwidth]{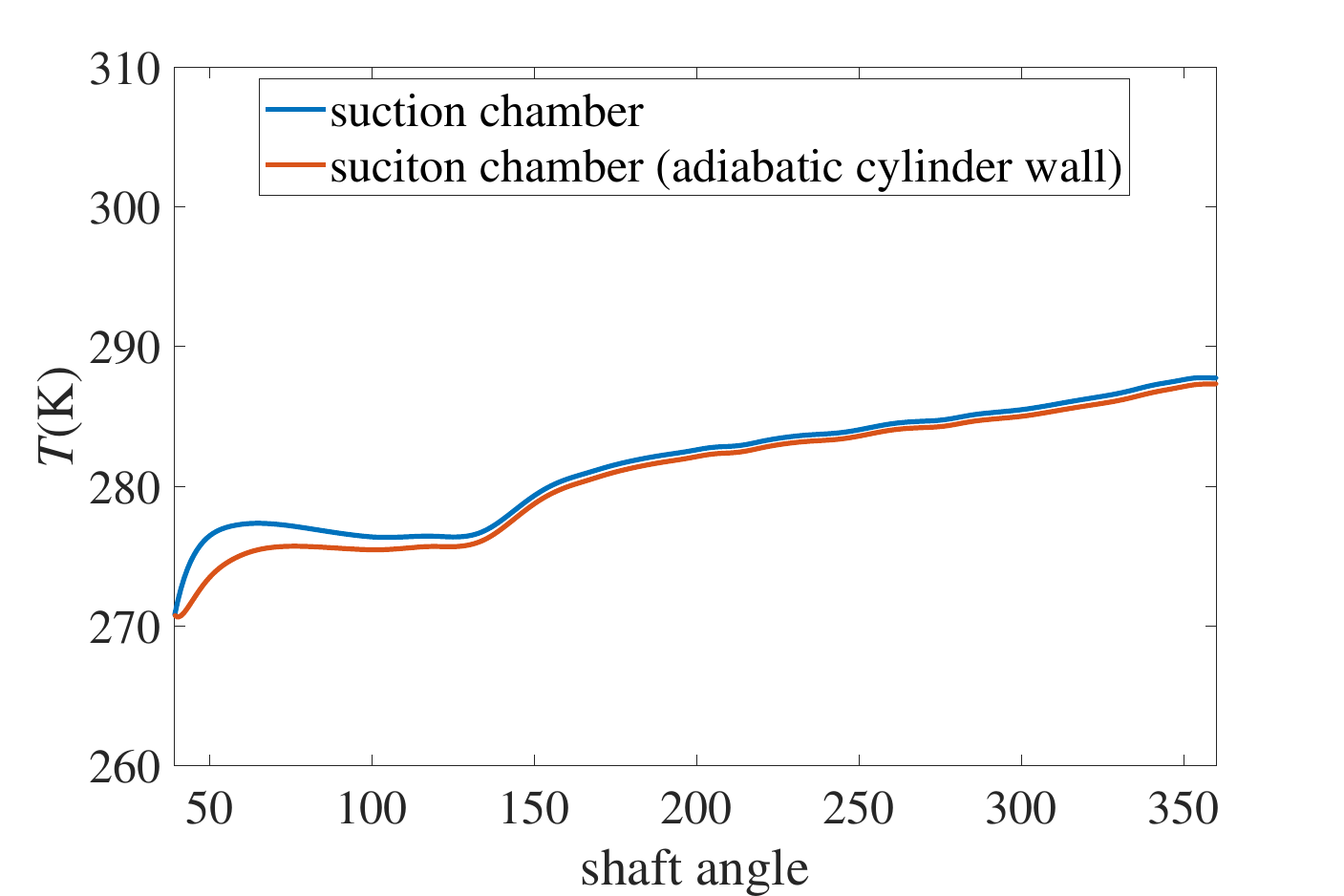}
    \caption{}
  \end{subfigure}             
  \begin{subfigure}[b]{0.49\textwidth}
    \includegraphics[width=\textwidth]{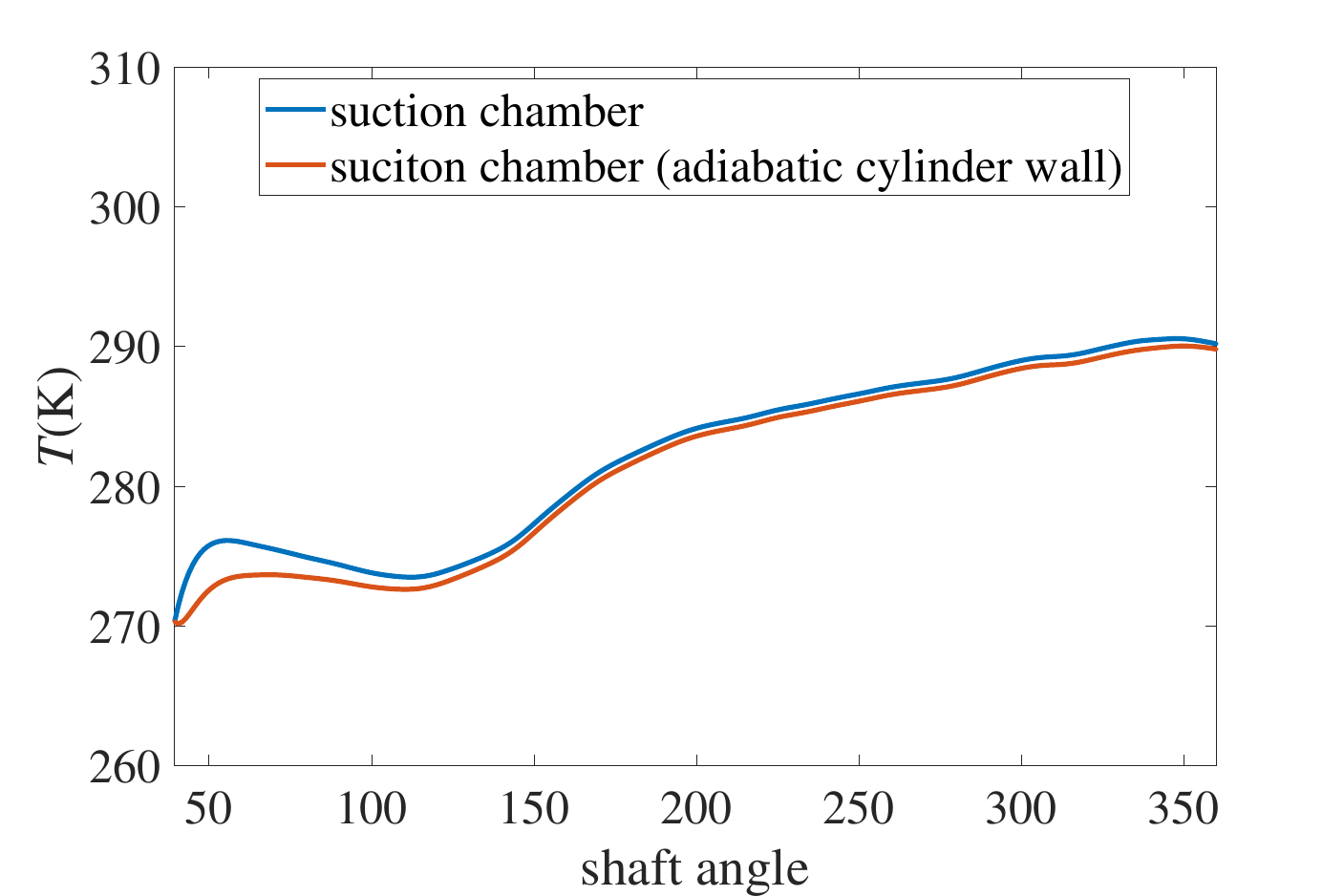}
    \caption{}
  \end{subfigure}             
  \begin{subfigure}[b]{0.49\textwidth}
    \includegraphics[width=\textwidth]{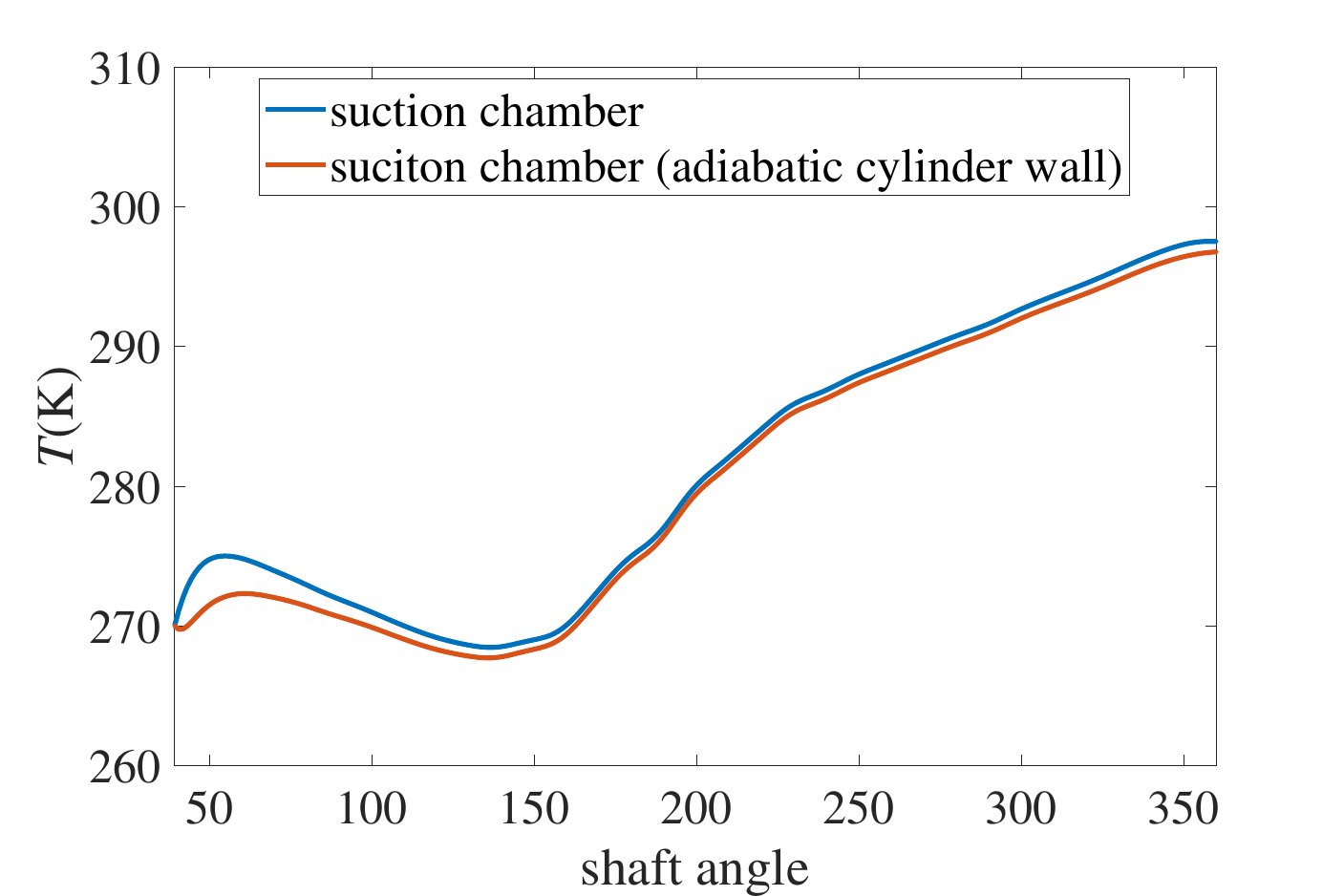}
    \caption{}
  \end{subfigure}             
  \begin{subfigure}[b]{0.49\textwidth}
    \includegraphics[width=\textwidth]{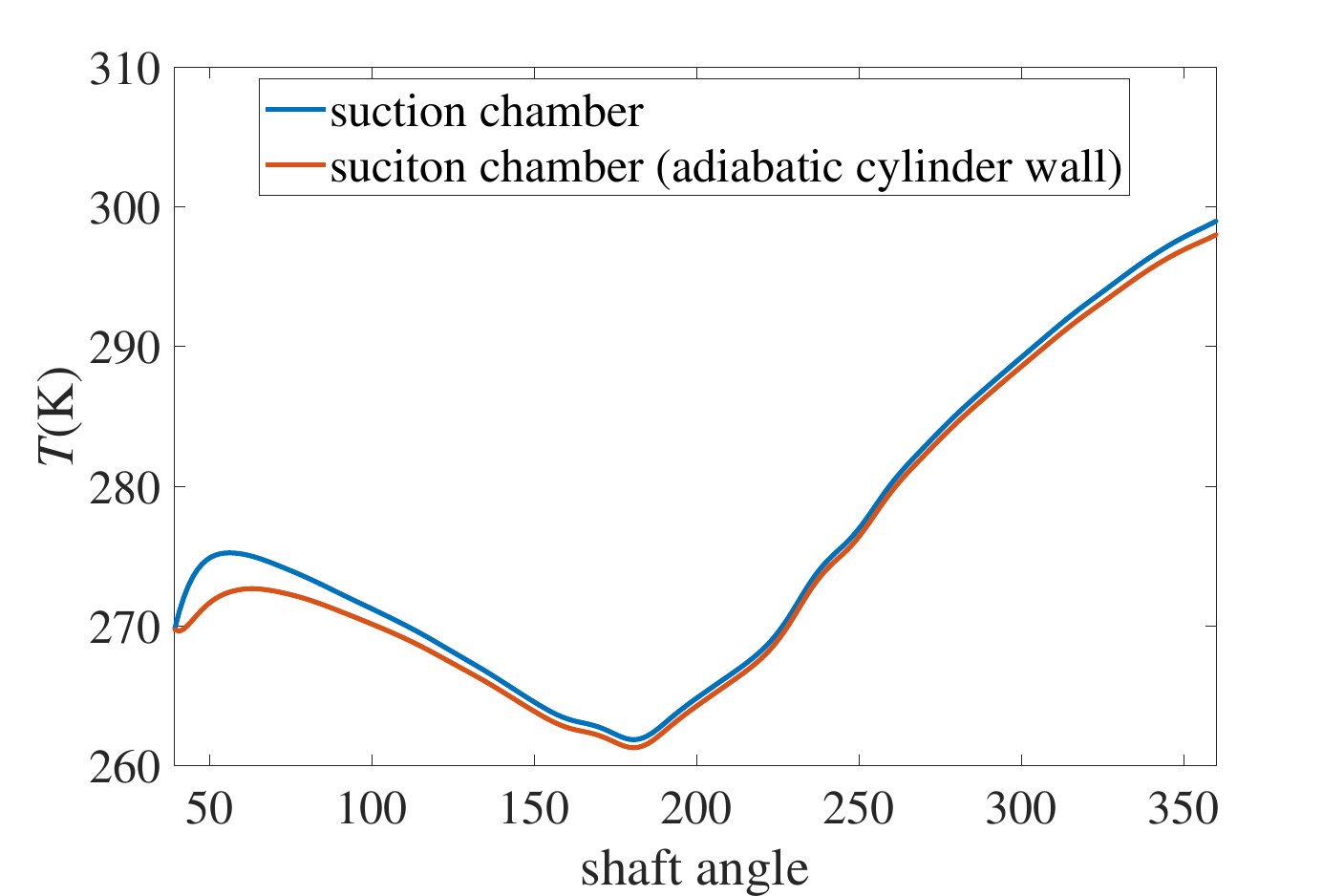}
    \caption{}
  \end{subfigure}             
  \caption{The temperature changes inside the suction chamber at (a) 80Hz, (b) 120Hz, (c) 140Hz, (d) 160Hz, (e) 180Hz, and (f) 200Hz.}
  \label{fig:Temperature}
\end{figure}

At high speeds, the supercharging effect causes losses in compressor efficiency for two reasons. First, the gas inside the suction pipe cannot fulfill the fast expansion of the suction chamber, causing an increased power loss in the suction process. Secondly, the temperature of the gas inside the suction chamber increases due to the supercharging effect at the end of the suction process. Figure \ref{fig:Temperature} illustrates the gas temperature inside the suction chamber during the suction process. To isolate the effect of supercharging on the gas heating, the calculation assumes an adiabatic cylinder wall (${d Q} = 0$ in equation \ref{eq:energy}). At speeds below 140Hz, the temperature of the refrigerant inside the suction chamber exhibits a relatively steady increase with the shaft angle. However, at higher speeds, there is a sudden drop to a minimum temperature, followed by a rapid increase to a maximum value at the end of the suction process. Notably, the temperature changes are proportional to the suction pressure changes. Comparison between the suction chamber with and without adiabatic conditions reveals that heat exchange between the cylinder wall and refrigerant gas only affects the temperature change during the initial suction process ($\alpha < 100^\circ$), indicating that pressure pulsation inside the suction chamber primarily determines the temperature change. At high speeds, the supercharging effect becomes significant, causing an increase in temperature inside the suction chamber at the end of the suction process. Consequently, the net effect of supercharging further reduces compressor efficiency at high speeds.

\newpage

\section{Conclusion} \label{sec:conclusion}

This study presents a comprehensive experimental investigation of compressor performance at high speeds. The experimental data validated the mathematical modeling of the volumetric efficiency, demonstrating the accuracy of the loss models proposed in this study.
By combining experimental data and mathematical modeling, the loss factors affecting the mass flow rate and power consumption were identified. The mathematical modeling of the volumetric efficiency was validated by the experimental data. The results indicate that the compressor efficiency degrades at speeds above 160Hz, with the volumetric efficiency increasing and the cylinder process efficiency decreasing.

The most prominent feature of the compressor operating at high speeds is the supercharging effect of significant pressure pulsation, which results in a great gain in volumetric efficiency. The major reasons for the volumetric efficiency loss are leakage and suction gas heating, which become more significant at higher speeds. Moreover, the measurement results on the discharge valve displacement indicate that the valve closes later at higher speeds, but the backflow loss is negligible due to the throttling effect.

The supercharging effect causes the most significant power loss due to the extra work done according to the volumetric efficiency gain. The discharge loss is caused by over-compression and the discharge gas pulsation, which is influenced by the fast compression process and valve inertia and stiffness. The suction power loss also becomes significant at speeds over 160Hz due to the gas inside the suction pipe can not fill the suction chamber fast enough during some portions of the suction process.

The net effects of each loss factor show that leakage loss and discharge loss contribute the most to efficiency degradation, which can be mitigated by optimizing the compressor's internal dimension. The supercharging effect is detrimental to the compressor efficiency, with its impact becoming significant at speeds above 140Hz due to the large pressure pulsation and temperature increase at the end of the suction process. With the key loss factors identified in this study, future work can be done to improve the compressor efficiency at high speeds.  

\section*{Acknowledgments}
We would like to thank the Midea Corporate Research Center for the support of this work. The study was conducted at Midea Global Innovative Center. (\url{https://www.midea.com.cn/About-Us/Innovation/research-centers})

\section*{Declarations}
The data that support the findings of this study are available from the corresponding author upon reasonable request.

%\bibliographystyle{elsarticle-num-names} \bibliography{cas-refs}

\begin{thebibliography}{26}
\expandafter\ifx\csname natexlab\endcsname\relax\def\natexlab#1{#1}\fi
\providecommand{\url}[1]{\texttt{#1}}
\providecommand{\href}[2]{#2}
\providecommand{\path}[1]{#1}
\providecommand{\DOIprefix}{doi:}
\providecommand{\ArXivprefix}{arXiv:}
\providecommand{\URLprefix}{URL: }
\providecommand{\Pubmedprefix}{pmid:}
\providecommand{\doi}[1]{\href{http://dx.doi.org/#1}{\path{#1}}}
\providecommand{\Pubmed}[1]{\href{pmid:#1}{\path{#1}}}
\providecommand{\bibinfo}[2]{#2}
\ifx\xfnm\relax \def\xfnm[#1]{\unskip,\space#1}\fi
%Type = Article
\bibitem[{Aw and Ooi(2021)}]{aw2021review}
\bibinfo{author}{K.~T. Aw}, \bibinfo{author}{K.~T. Ooi},
\newblock \bibinfo{title}{A review on sliding vane and rolling piston compressors},
\newblock \bibinfo{journal}{Machines} \bibinfo{volume}{9} (\bibinfo{year}{2021}) \bibinfo{pages}{125}.
%Type = Article
\bibitem[{Wu et~al.(2015)Wu, Wang, Li, Chen, Gao, Chen, and Jiang}]{wu2015experimental}
\bibinfo{author}{J.~Wu}, \bibinfo{author}{G.~Wang}, \bibinfo{author}{Y.~Li}, \bibinfo{author}{A.~Chen}, \bibinfo{author}{Q.~Gao}, \bibinfo{author}{Z.~Chen}, \bibinfo{author}{B.~Jiang},
\newblock \bibinfo{title}{Experimental study on pv diagram and valve displacement of a hc290 rotary compressor},
\newblock \bibinfo{journal}{Proceedings of the Institution of Mechanical Engineers, Part C: Journal of Mechanical Engineering Science} \bibinfo{volume}{229} (\bibinfo{year}{2015}) \bibinfo{pages}{3113--3124}.
%Type = Article
\bibitem[{Wakabayashi et~al.(1982)Wakabayashi, Yuuda, Aizawa, and Yamamura}]{wakabayashi1982analysis}
\bibinfo{author}{H.~Wakabayashi}, \bibinfo{author}{J.~Yuuda}, \bibinfo{author}{T.~Aizawa}, \bibinfo{author}{M.~Yamamura},
\newblock \bibinfo{title}{Analysis of performance in a rotary compressor},
\newblock \bibinfo{journal}{International compressor engineering conference}  (\bibinfo{year}{1982}).
%Type = Article
\bibitem[{Nagatomo et~al.(1984)Nagatomo, Sakata, Tago, and Hattori}]{nagatomo1984performance}
\bibinfo{author}{S.~Nagatomo}, \bibinfo{author}{H.~Sakata}, \bibinfo{author}{M.~Tago}, \bibinfo{author}{H.~Hattori},
\newblock \bibinfo{title}{Performance analysis of rolling piston type rotary compressor for household refrigerators},
\newblock \bibinfo{journal}{International compressor engineering conference}  (\bibinfo{year}{1984}).
%Type = Article
\bibitem[{Monasry et~al.(2018)Monasry, Hirayama, Aoki, Shida, Hatayama, and Okada}]{monasry2018development}
\bibinfo{author}{J.~F. Monasry}, \bibinfo{author}{T.~Hirayama}, \bibinfo{author}{T.~Aoki}, \bibinfo{author}{S.~Shida}, \bibinfo{author}{M.~Hatayama}, \bibinfo{author}{M.~Okada},
\newblock \bibinfo{title}{Development of large capacity and high efficiency rotary compressor},
\newblock \bibinfo{journal}{International compressor engineering conference}  (\bibinfo{year}{2018}).
%Type = Article
\bibitem[{Matsuzaka and Nagatomo(1982)}]{matsuzaka1982rolling}
\bibinfo{author}{T.~Matsuzaka}, \bibinfo{author}{S.~Nagatomo},
\newblock \bibinfo{title}{Rolling piston type rotary compressor performance analysis},
\newblock \bibinfo{journal}{International compressor engineering conference}  (\bibinfo{year}{1982}).
%Type = Article
\bibitem[{Liu and Soedel(1994)}]{liu1994performance}
\bibinfo{author}{Z.~Liu}, \bibinfo{author}{W.~Soedel},
\newblock \bibinfo{title}{Performance study of a variable speed compressor with special attention to supercharging effect},
\newblock \bibinfo{journal}{International compressor engineering conference}  (\bibinfo{year}{1994}).
%Type = Article
\bibitem[{Youn et~al.(1998)Youn, Park, Bae, and Ma}]{youn1998design}
\bibinfo{author}{Y.~Youn}, \bibinfo{author}{S.~Park}, \bibinfo{author}{J.~Bae}, \bibinfo{author}{Y.~Ma},
\newblock \bibinfo{title}{Design optimization for the discharge system of the rotary compressor using alternative refrigerant r410a},
\newblock \bibinfo{journal}{International compressor engineering conference}  (\bibinfo{year}{1998}).
%Type = Article
\bibitem[{Lee et~al.(2012)Lee, Bai, and Shim}]{lee2012performance}
\bibinfo{author}{S.~Lee}, \bibinfo{author}{C.~Bai}, \bibinfo{author}{J.~Shim},
\newblock \bibinfo{title}{Performance analysis and experiment of new 3d rotary compressor},
\newblock \bibinfo{journal}{Proceedings of the Institution of Mechanical Engineers, Part C: Journal of Mechanical Engineering Science} \bibinfo{volume}{226} (\bibinfo{year}{2012}) \bibinfo{pages}{133--144}.
%Type = Article
\bibitem[{Nomura et~al.(1984)Nomura, Ohta, Takeshita, and Ozawa}]{nomura1984efficiency}
\bibinfo{author}{T.~Nomura}, \bibinfo{author}{M.~Ohta}, \bibinfo{author}{K.~Takeshita}, \bibinfo{author}{Y.~Ozawa},
\newblock \bibinfo{title}{Efficiency improvement in rotary compressor},
\newblock \bibinfo{journal}{International compressor engineering conference}  (\bibinfo{year}{1984}).
%Type = Article
\bibitem[{Kim et~al.(2017)Kim, Min, Na, Choi, and Kim}]{kim2017estimation}
\bibinfo{author}{G.~Kim}, \bibinfo{author}{B.~Min}, \bibinfo{author}{S.~Na}, \bibinfo{author}{G.~Choi}, \bibinfo{author}{D.~Kim},
\newblock \bibinfo{title}{Estimation of leakage through radial clearance during compression process of a rolling piston rotary compressor},
\newblock \bibinfo{journal}{Journal of Mechanical Science and Technology} \bibinfo{volume}{31} (\bibinfo{year}{2017}) \bibinfo{pages}{6033--6040}.
%Type = Article
\bibitem[{Ba et~al.(2016)Ba, Deng, Che, Li, Guo, Li, and Yue}]{ba2016gas}
\bibinfo{author}{D.-C. Ba}, \bibinfo{author}{W.-J. Deng}, \bibinfo{author}{S.-G. Che}, \bibinfo{author}{Y.~Li}, \bibinfo{author}{H.-X. Guo}, \bibinfo{author}{N.~Li}, \bibinfo{author}{X.-J. Yue},
\newblock \bibinfo{title}{Gas dynamics analysis of a rotary compressor based on cfd},
\newblock \bibinfo{journal}{Applied Thermal Engineering} \bibinfo{volume}{99} (\bibinfo{year}{2016}) \bibinfo{pages}{1263--1269}.
%Type = Article
\bibitem[{Wu and Wang(2013)}]{wu2013numerical}
\bibinfo{author}{J.~Wu}, \bibinfo{author}{G.~Wang},
\newblock \bibinfo{title}{Numerical study on oil supply system of a rotary compressor},
\newblock \bibinfo{journal}{Applied thermal engineering} \bibinfo{volume}{61} (\bibinfo{year}{2013}) \bibinfo{pages}{425--432}.
%Type = Article
\bibitem[{Cipollone et~al.(2006)Cipollone, Contaldi, Sciarretta, Tufano, and Villante}]{cipollone2006theoretical}
\bibinfo{author}{R.~Cipollone}, \bibinfo{author}{G.~Contaldi}, \bibinfo{author}{A.~Sciarretta}, \bibinfo{author}{R.~Tufano}, \bibinfo{author}{C.~Villante},
\newblock \bibinfo{title}{A theoretical model and experimental validation of a sliding vane rotary compressor},
\newblock \bibinfo{journal}{International compressor engineering conference}  (\bibinfo{year}{2006}).
%Type = Article
\bibitem[{Wu(2000)}]{wu2000mathematical}
\bibinfo{author}{J.~Wu},
\newblock \bibinfo{title}{A mathematical model for internal leakage in a rotary compressor},
\newblock \bibinfo{journal}{International compressor engineering conference}  (\bibinfo{year}{2000}).
%Type = Article
\bibitem[{Mathison et~al.(2008)Mathison, Braun, and Groll}]{mathison2008modeling}
\bibinfo{author}{M.~M. Mathison}, \bibinfo{author}{J.~E. Braun}, \bibinfo{author}{E.~A. Groll},
\newblock \bibinfo{title}{Modeling of a two-stage rotary compressor},
\newblock \bibinfo{journal}{HVAC\&R Research} \bibinfo{volume}{14} (\bibinfo{year}{2008}) \bibinfo{pages}{719--748}.
%Type = Article
\bibitem[{Cai et~al.(2015)Cai, He, Yokoyama, Tian, Yang, and Pan}]{cai2015simulation}
\bibinfo{author}{D.~Cai}, \bibinfo{author}{G.~He}, \bibinfo{author}{T.~Yokoyama}, \bibinfo{author}{Q.~Tian}, \bibinfo{author}{X.~Yang}, \bibinfo{author}{J.~Pan},
\newblock \bibinfo{title}{Simulation and comparison of leakage characteristics of r290 in rolling piston type rotary compressor},
\newblock \bibinfo{journal}{International Journal of Refrigeration} \bibinfo{volume}{53} (\bibinfo{year}{2015}) \bibinfo{pages}{42--54}.
%Type = Article
\bibitem[{Padhy and Dwivedi(1994)}]{padhy1994heat}
\bibinfo{author}{S.~K. Padhy}, \bibinfo{author}{S.~N. Dwivedi},
\newblock \bibinfo{title}{Heat transfer analysis of a rolling-piston rotary compressor},
\newblock \bibinfo{journal}{International journal of refrigeration} \bibinfo{volume}{17} (\bibinfo{year}{1994}) \bibinfo{pages}{400--410}.
%Type = Article
\bibitem[{Liu and Soedel(1992)}]{liu1992modeling}
\bibinfo{author}{Z.~Liu}, \bibinfo{author}{W.~Soedel},
\newblock \bibinfo{title}{Modeling temperatures in high speed compressors for the purpose of gas pulsation and valve loss modelling},
\newblock \bibinfo{journal}{International compressor engineering conference}  (\bibinfo{year}{1992}).
%Type = Book
\bibitem[{Liu(1993)}]{liu1993simulation}
\bibinfo{author}{Z.~Liu}, \bibinfo{title}{Simulation of a variable speed compressor with special attention to supercharging effects}, \bibinfo{publisher}{Purdue e-Pubs}, \bibinfo{year}{1993}.
%Type = Book
\bibitem[{Mills(1992)}]{mills1992heat}
\bibinfo{author}{A.~F. Mills}, \bibinfo{title}{Heat transfer}, \bibinfo{publisher}{CRC Press}, \bibinfo{year}{1992}.
%Type = Article
\bibitem[{YANAGISAWA et~al.(1984)YANAGISAWA, SHIMIZU, DOHI, and NIHASHI}]{1984741}
\bibinfo{author}{T.~YANAGISAWA}, \bibinfo{author}{T.~SHIMIZU}, \bibinfo{author}{M.~DOHI}, \bibinfo{author}{S.~NIHASHI},
\newblock \bibinfo{title}{A study on suction gas heating in a rolling piston rotary compressor},
\newblock \bibinfo{journal}{Bulletin of JSME} \bibinfo{volume}{27} (\bibinfo{year}{1984}) \bibinfo{pages}{741--748}.
%Type = Article
\bibitem[{Pandeya and Soedel(1978)}]{pandeya1978rolling}
\bibinfo{author}{P.~Pandeya}, \bibinfo{author}{W.~Soedel},
\newblock \bibinfo{title}{Rolling piston type rotary compressors with special attention to friction and leakage},
\newblock \bibinfo{journal}{International compressor engineering conference}  (\bibinfo{year}{1978}).
%Type = Article
\bibitem[{Lenz and Cooksey(1994)}]{lenz1994application}
\bibinfo{author}{J.~R. Lenz}, \bibinfo{author}{E.~Cooksey},
\newblock \bibinfo{title}{Application of computational fluid dynamics to compressor efficiency improvement},
\newblock \bibinfo{journal}{International compressor engineering conference}  (\bibinfo{year}{1994}).
%Type = Article
\bibitem[{Yanagisawa and Shimizu(1985)}]{yanagisawa1985leakage}
\bibinfo{author}{T.~Yanagisawa}, \bibinfo{author}{T.~Shimizu},
\newblock \bibinfo{title}{Leakage losses with a rolling piston type rotary compressor. i. radical clearance on the rolling piston},
\newblock \bibinfo{journal}{International Journal of Refrigeration} \bibinfo{volume}{8} (\bibinfo{year}{1985}) \bibinfo{pages}{75--84}.
%Type = Article
\bibitem[{Prater~Jr and Hnat(2003)}]{prater2003optical}
\bibinfo{author}{G.~Prater~Jr}, \bibinfo{author}{W.~P. Hnat},
\newblock \bibinfo{title}{Optical measurement of discharge valve modal parameters for a rolling piston refrigeration compressor},
\newblock \bibinfo{journal}{Measurement} \bibinfo{volume}{33} (\bibinfo{year}{2003}) \bibinfo{pages}{75--84}.

\end{thebibliography}





% \end{thebibliography}
\end{document}